\documentclass[twocolumn,british,aps,superscriptaddress,prc,nofootinbib,showpacs,showkeys]{revtex4-1}
\usepackage[T1]{fontenc}
\usepackage[latin9]{inputenc}
\usepackage{color}
\usepackage{babel}
\usepackage{amsmath}
\usepackage{graphicx}
\usepackage{esint}
\usepackage[unicode=true,pdfusetitle,
 bookmarks=true,bookmarksnumbered=false,bookmarksopen=false,
 breaklinks=false,pdfborder={0 0 1},backref=false,colorlinks=true]
 {hyperref}
\hypersetup{
 pdfborderstyle=,linkcolor=blue,citecolor=cyan}

\makeatletter
\usepackage{babel}

\usepackage{xcolor}

\makeatother

\begin{document}
\title{Electromagnetic anomaly in the presence of electric and chiral magnetic
conductivities in relativistic heavy-ion collisions}
\author{Irfan Siddique}
\email{irfansiddique@sdu.edu.cn}

\affiliation{Institute of Frontier and Interdisciplinary Science, Shandong University,
Qingdao, Shandong, 266237, China}
\author{Shanshan Cao}
\email{shanshan.cao@sdu.edu.cn}

\affiliation{Institute of Frontier and Interdisciplinary Science, Shandong University,
Qingdao, Shandong, 266237, China}
\author{Uzma Tabassam}
\affiliation{Department of Physics, COMSATS University Islamabad Campus, Islamabad,
Park Road, 44000, Pakistan}
\author{Mohsin Saeed}
\affiliation{Department of Physics, University of the Punjab, Quaid-e-Azam Campus,
Lahore, 54590, Pakistan}
\author{Muhammad Waqas}
\affiliation{School of Nuclear Science and Technology, University of Chinese Academy
of Sciences, Beijing, 100049, China}
\begin{abstract}
We study the spacetime evolution of electric (\textbf{E}) and magnetic
(\textbf{B}) fields along with the electromagnetic anomaly \textbf{$\text{\ensuremath{\left(\textbf{E}\cdot\textbf{B}\right)}}$}
in the presence of electric ($\sigma$) and chiral magnetic ($\sigma_{\chi}$)
conductivities in Au+Au collisions at $\sqrt{s_{\mathrm{NN}}}=200$~GeV.
By comparing to the Lienard-Wiechert solutions with zero conductivities,
we observe a symmetry breaking of the electromagnetic field in a conducting
medium with respect to the reaction plane. The decay of the field
is also significantly decelerated after the conductivities are introduced.
Similar effects are also found for the dipole structure of $\textbf{E}\cdot\textbf{B}$
as well as the quadrupole structure of $(\textbf{E}\cdot\textbf{B})\textbf{B}$,
which may finally affect the charge separation of the elliptic flow
coefficient of hadrons observed in high-energy nuclear collisions.
\end{abstract}
\maketitle

\section{Introduction}

The high-speed movement of charged nuclei in non-central relativistic
heavy-ion collisions can produce strong electromagnetic field. 
Its magnitude can be estimated via $eB\sim\gamma vZe^{2}/R_{A}^{2}$,
whose peak value can reach the order of $10^{18}$~Gauss in Au+Au
collisions at the BNL Relativistic Heavy-Ion Collider (RHIC), and
$10^{19}$~Gauss in Pb+Pb collisions at the CERN Large Hadron Collider
(LHC)~\citep{Bzdak:2011yy,Deng:2012pc,Zhao:2017nfq,Zhong:2014cda}.
This provides a unique environment to investigate properties of nuclear
matter under extreme electromagnetic field, such as the anomalous
transport effects in the quark-gluon plasma (QGP) produced by the
energetic nuclear collisions~\citep{Kharzeev:2015znc,Kharzeev:2007tn,Kharzeev:2007jp,Fukushima:2008xe,Liao:2014ava,Huang:2013iia}.
The electromagnetic field can also cause separation of particles with
opposite charges, as reflected by the charge-odd directed flow coefficient
found in both theoretical calculations~\citep{Gursoy:2014aka,Gursoy:2018yai,Chatterjee:2018lsx,Inghirami:2019mkc,Oliva:2020mfr,Sun:2021psy,Zhang:2022lje}
and experimental measurements~\citep{STAR:2019clv,ALICE:2019sgg},
although a precise agreement between theory and experiment is still an ongoing effort.

With the presence of an external electric field $\textbf{E}$, one
would expect a vector current induced in a conducting matter according
to the Ohm's law, 
\begin{equation}
\textbf{j}_{V}=\sigma\textbf{E},
\end{equation}
with $\sigma$ being the electric conductivity and $\textbf{j}_{V}$
being the electric current. Meanwhile, inside a plasma composed of
chiral fermions, vector current $j_{V}^{\mu}$ and axial current $j_{A}^{\mu}$
can also be induced by magnetic field $\textbf{B}$. 
If chiral anomaly, or a nonzero axial chiral charge potential $\mu_{A}$,
exists, a vector current $\textbf{j}_{V}=\left\langle \overline{\psi}\gamma^{i}\psi\right\rangle $
will be induced by the imbalance between left and right handed quarks
as~\citep{Son:2009tf,Kharzeev:2010gd,Burnier:2011bf}: 
\begin{equation}
\textbf{j}_{V}^{\mathrm{CME}}=\sigma_{5}\mu_{A}\textbf{B},\label{eq:eq0_1}
\end{equation}
where $\sigma_{5}$ is known as the chiral magnetic conductivity given
by $\sigma_{5}\simeq eN_{c}/2\pi^{2}$ with $N_{c}$ being the number
of colors. This is known as the chiral magnetic effect (CME). With
the mass term included, the axial anomaly equation reads~\citep{Guo:2016nnq,Iatrakis:2015fma},
\begin{align}
\partial_{\mu}j_{5}^{\mu} & =2im\overline{\psi}\gamma^{5}\psi-\frac{e^{2}}{16\pi^{2}}\epsilon^{\mu\nu\rho\sigma}F_{\mu\nu}F_{\rho\sigma}\nonumber \\
 & \;\;\;\;-\frac{g^{2}}{16\pi^{2}}\text{tr}\epsilon^{\mu\nu\rho\sigma}G_{\mu\nu}G_{\rho\sigma},
\end{align}
from which one can observe three terms of contributions to the axial
charge. The first term on the right hand side comes from the finite
quark mass originated from the chiral symmetry breaking, while the
second and third terms correspond to the QED and QCD anomalies respectively.
The QED anomaly, usually represented by $\textbf{E}\cdot\textbf{B}$,
will be the main focus of this work. 

Similarly, a nonzero vector chemical potential $\mu_{A}$ would induce
an axial current $\textbf{j}_{A}=\left\langle \overline{\psi}\gamma^{i}\gamma^{5}\psi\right\rangle $
in the presence of the external $\textbf{B}$ field~\citep{Metlitski:2005pr,Son:2004tq}:
\begin{equation}
\textbf{j}_{A}^{\mathrm{CSE}}=\sigma_{5}\mu_{V}\textbf{B},
\end{equation}
which causes the axial charge separation along the $\textbf{B}$ and
is known as the chiral separation effect (CSE).

Chiral magnetic and chiral separation effects have been investigated
within various approaches, such as hydrodynamics, kinetic theory,
holographic QCD and lattice QCD~\citep{Son:2012wh,Stephanov:2012ki,Gao:2012ix,Chen:2012ca,Chen:2013iga,Chen:2013tra,Satow:2014lva}.
It has been proposed that the nonzero vector and axial charges can
mutually induce each other, leading to a collective excitation in
the QGP known as the chiral magnetic wave (CMW)~\citep{Kharzeev:2010gd,Burnier:2012ae,Yee:2013cya}.
The charge quadrupole structure associated with this CMW can further
result in different elliptic flow coefficients ($v_{2}$) between
positive and negative charged particles~\cite{Burnier:2011bf,Ma:2014iva},
as observed by the STAR experiment~\cite{STAR:2015wza}. Nevertheless,
it has been suggested in Ref.~\cite{Zhao:2019ybo} that even without
the formation of CMW, the dipole shape of the $\textbf{E}\cdot\textbf{B}$
distribution in the transverse plane can already generate an electric
quadrupole moment when being coupled to the magnetic field $\textbf{B}$.
This provides an alternative direction for understanding the $v_{2}$
separation between opposite charges, considering the negative result
on the recent search for the CME at RHIC~\cite{STAR:2021mii}. Moreover,
this novel mechanism does not need a finite baryon density ($\mu_{V}$)
to drive the charge separation of $v_{2}$, therefore may lead to
different beam energy dependence of this charge separation, which
can be further tested by the beam energy scan program at RHIC. The
space-averaged electromagnetic field and electromagnetic anomaly weighted
by energy density have been further investigated in Ref.~\cite{Siddique:2021smf}.

In this work, we extend these previous studies~\cite{Zhao:2019ybo,Siddique:2021smf}
on the spatial distribution of $\textbf{E}\cdot\textbf{B}$ to a more
realistic nuclear medium that includes both electric and chiral magnetic
conductivities. It has been found that the decay of the electromagnetic
field can be significantly decelerated when conductivities are introduced~\citep{Tuchin:2013apa,Tuchin:2014iua,Li:2016tel,Chen:2021nxs,McLerran:2013hla,Gursoy:2014aka,Inghirami:2016iru}.
Symmetry breaking has also been suggested for the field with respect
to the reaction plane after including the conductivities~\cite{Li:2016tel}.
We will follow Ref.~\cite{Li:2016tel} to further investigate the
time evolution of the spatial distribution of the electromagnetic
field in the presence of electric and chiral magnetic conductivities.
In particular, effects on the dipole structure of the electromagnetic
anomaly and the electric quadrupole pattern will be discussed in detail.

This work will be organized as follows. We will first provide a brief
review on the solution of the electromagnetic field in both conducting
and non-conducting media in Sec.~\ref{sec:2}. Numerical results
of the spacetime evolution of the electromagnetic field will be presented
in Sec.~\ref{sec:3}, and compared between with and without including
electric and chiral magnetic conductivities. In Sec.~\ref{sec:4},
we will discuss effects of conductivities on the electromagnetic anomaly
and the electric quadrupole moment. A summary and outlook will be
presented in Sec.~\ref{sec:summary}.


\section{\label{sec:2} Calculation of electromagnetic field}

\subsection*{A: Non-conducting System ($\sigma=\sigma_{\chi}=0$)}

For a non-conducting system, or vacuum, where both electric and chiral
magnetic conductivities are zero ($\sigma=\sigma_{\chi}=0$), we evaluate
the electromagnetic field according to the Lienard-Wiechert potential~\citep{Bzdak:2011yy,Bloczynski:2012en}
as 
\begin{equation}
{\textbf{E}}(t,\textbf{x})=\frac{e}{4\pi}\sum_{n}\frac{\left(1-v_{n}^{2}\right)\textbf{R}_{n}}{\left(\textbf{R}_{n}^{2}-\left(\textbf{R}_{n}\times\textbf{v}_{n}\right)^{2}\right)^{3/2}},\label{eq:LW_E}
\end{equation}
\begin{equation}
{\textbf{B}}(t,\textbf{x})=\frac{e}{4\pi}\sum_{n}\frac{\left(1-v_{n}^{2}\right)\left(\textbf{v}_{n}\times\textbf{R}_{n}\right)}{\left(\textbf{R}_{n}^{2}-\left(\textbf{R}_{n}\times\textbf{v}_{n}\right)^{2}\right)^{3/2}},\label{eq:LW_B}
\end{equation}
where $\textbf{R}_{n}=\textbf{x}-\textbf{x}_{n}$ is the relative
position vector between the field point $\textbf{x}$ under discussion
and the source point $\textbf{x}_{n}$, and $\textbf{x}_{n}$ and
$\textbf{v}_{n}$ are respectively the position and velocity of the
$n$-th proton in the colliding nuclei at the current time $t$. Note
that the above equations are valid when each source charge is traveling
with a constant velocity. Otherwise, the original form of the Lienard-Wiechert
fields~\cite{Deng:2012pc} using the retarded time should be applied.

\subsection*{B: Conducting System ($\sigma\protect\neq0,\sigma_{\chi}\protect\neq0$)}

The QGP matter produced in heavy-ion collisions is a conducting medium.
The in-medium electromagnetic field can be solved using the Maxwell
equations with both electric ($\sigma$) and chiral magnetic ($\sigma_{\chi}$)
conductivities included: 
\begin{eqnarray}
\nabla\cdot\textbf{F} & = & \left\{ \begin{array}{cc}
\rho_{\mathrm{ext}}/\epsilon & \rightarrow\textbf{E}\\
0 & \rightarrow\textbf{B}
\end{array}\right.,\label{eq:eq1}\\
\nabla\times\textbf{F} & = & \left\{ \begin{array}{cc}
-\partial_{t}\textbf{B}\text{ \ \ \ \ \ \ \ \ \ \ } & \rightarrow\textbf{E}\\
\partial_{t}\textbf{E}+\textbf{J}_{\mathrm{ext}}+\sigma\textbf{E}+\sigma_{\chi}\textbf{B} & \rightarrow\textbf{B}
\end{array}\right.,\label{eq:eq1.2}
\end{eqnarray}
where $\rho_{\mathrm{ext}}$ and $\textbf{J}_{\mathrm{ext}}$ are
external charge and current densities, and $\textbf{F}$ denotes either
electric ($\textbf{E}$) or magnetic ($\textbf{B}$) field. Considering
that all source charges propagate along the $z$-axis, one can obtain
the following algebraic solutions of the electromagnetic field using
the Green's function method in the cylindrical coordinates~\citep{Li:2016tel}:
\begin{align}
B_{\phi}(t,\textbf{x}) & =\frac{Q}{4\pi}\frac{v\gamma x_{\mathrm{T}}}{\Delta^{3/2}}\left(1+\frac{\sigma v\gamma}{2}\sqrt{\Delta}\right)e^{A},\nonumber \\
B_{r}(t,\textbf{x}) & =-\sigma_{\chi}\frac{Q}{8\pi}\frac{v\gamma^{2}x_{\mathrm{T}}}{\Delta^{3/2}}e^{A}\left[\gamma\left(vt-z\right)+A\sqrt{\Delta}\right],\nonumber \\
B_{z}(t,\textbf{x}) & =\sigma_{\chi}\frac{Q}{8\pi}\frac{v\gamma}{\Delta^{3/2}}e^{A}\Big[\Delta\left(1-\frac{\sigma v\gamma}{2}\sqrt{\Delta}\right)\nonumber \\
 & \;\;\;\;+\gamma^{2}\left(vt-z\right)^{2}\left(1+\frac{\sigma v\gamma}{2}\sqrt{\Delta}\right)\Big],\label{eq:eq8}
\end{align}
in which $\Delta$ and $A$ are defined as $\Delta\equiv\gamma^{2}\left(vt-z\right)^{2}+x_{\mathrm{T}}^{2}$
and $A\equiv\left(\sigma v\gamma/2\right)\left[\gamma\left(vt-z\right)-\sqrt{\Delta}\right]$, with $x_{\mathrm{T}}$ being the magnitude of the transverse coordinate $x_{\mathrm{T}}=\sqrt{x^{2}+y^{2}}$; and 
\begin{align}
E_{\phi}(t,\textbf{x}) & =\sigma_{\chi}\frac{Q}{8\pi}\frac{v^{2}\gamma^{2}x_{\mathrm{T}}}{\Delta^{3/2}}e^{A}\left[\gamma\left(vt-z\right)+A\sqrt{\Delta}\right],\nonumber \\
E_{r}(t,\textbf{x}) & =\frac{Q}{4\pi}e^{A}\Bigg\{\frac{\gamma x_{\mathrm{T}}}{\Delta^{3/2}}\left(1+\frac{\sigma v\gamma}{2}\sqrt{\Delta}\right)\nonumber \\
 & \;\;\;\;-\frac{\sigma}{vx_{\mathrm{T}}}e^{-\sigma(t-z/v)}\left[1+\frac{\gamma\left(vt-z\right)}{\sqrt{\Delta}}\right]\Bigg\},\nonumber \\
E_{z}(t,\textbf{x}) & =\frac{Q}{4\pi}\left\{ -\frac{e^{A}}{\Delta^{3/2}}\left[\gamma\left(vt-z\right)+A\sqrt{\Delta}+\frac{\sigma\gamma}{v}\Delta\right]\right.\nonumber \\
 & \;\;\;\;\left.+\frac{\sigma^{2}}{v^{2}}e^{-\sigma\left(t-z/v\right)}\Gamma\left(0,-A\right)\right\} ,\label{eq:eq9}
\end{align}
with $\Gamma\left(0,-A\right)$ being the incomplete gamma function
defined as $\Gamma\left(a,z\right)=\int_{z}^{\infty}t^{a-1}\exp\left(-t\right)\,dt$.
One may verify that Eqs.~(\ref{eq:eq8}) and~(\ref{eq:eq9}) above
return to the previous Lienard-Wiechert solution with vanishing $\sigma$
and $\sigma_{\chi}$.

\begin{figure*}
\begin{centering}
\includegraphics[width=10cm]{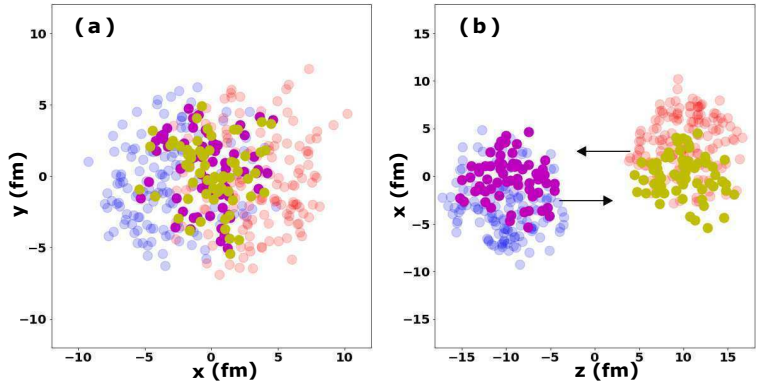} 
\par\end{centering}
\caption{\label{fig:3.1Au+Au-initial-geometry} (Color online) Initial geometry
of a Au+Au collision event generated by the MC Glauber model for $b=10$~fm
and $\sqrt{s_{\mathrm{NN}}}=200$~GeV, left for the view on the transverse
plane and right for the view on the reaction plane.}
\end{figure*}

In this work, we use the Monte-Carlo (MC) Glauber model developed
by the PHOBOS Collaboration~\citep{Loizides:2014vua} to calculate
the spatial distribution of the source charges. A two-step calculation
is performed in this model. First, for a given impact parameter $b$,
the centers of projectile and target nuclei are located at $x=\pm b/2$
with 
the impact parameter defined in the $x$-direction and the beams in
the $z$-direction. The positions of nucleons in the two nuclei are
determined stochastically. 
The Woods-Saxon distribution is taken for the density profile of nucleons
in each nucleus 
\begin{equation}
\rho(r,\theta)=\frac{\rho_{0}}{1+\exp\left[\frac{r-R(\theta)}{d}\right]}\left[1+w\frac{r^{2}}{R(\theta)^{2}}\right],
\end{equation}
where $\rho_{0}$ denotes the nuclear density at the nucleus center,
$d$ is the surface thickness parameter, and $R(\theta)=R_{0}[1+\beta_{2}Y_{20}(\theta)+\beta_{4}Y_{40}(\theta)]$
is the nuclear radius in which $Y_{nl}(\theta)$ are spherical harmonic
functions. Here, the parameters $\beta_{2}$, $\beta_{4}$ and $w$
determine the deviation from a spherical nucleus. These nucleons are
then assumed to propagate along straight trajectories (in $+$/$-z$
directions). For each pair of nucleons, one from projectile and one
from target, a collision between them takes place if their distance
$d$ (in the transverse plane) satisfies $d\leq\sqrt{\sigma_{\mathrm{inel}}^{\mathrm{NN}}/\pi}$,
where $\sigma_{\mathrm{inel}}^{\mathrm{NN}}$ is the inelastic cross
section of nucleon-nucleon collisions. Those nucleons that participate
in collisions are labelled as ``participants'' while those that
do not participate in collisions are labelled as ``spectators''.


In this work, we use $\sigma_{\mathrm{inel}}^{\mathrm{NN}}=42\,\text{mb}$, $\rho_0=0.17$~fm$^{-3}$,
$R=6.38$~fm, $d=0.535$~fm and $w=0$ for Au-Au collisions at $\sqrt{s_{\mathrm{NN}}}=200$~GeV
at RHIC. We define the initial time ($t=0$) as the moment when the two oppositely moving nuclei collide. In Fig.~\ref{fig:3.1Au+Au-initial-geometry}, we illustrate
our initial charge distribution based on this MC Glauber approach,
left for the view on the transverse plane, and right for the view
on the reaction plane. Purple and green dots represent participant
nucleons from the two colliding nuclei, while blue and red represent
spectators that do not participate in inelastic scatterings. For each
nucleon, we use the probability $Z/A$ ($79/197$ for Au nucleus)
to determine whether it is a proton that contributes to the electromagnetic
field we discuss. Protons in both participants and spectators are
taken into account for evaluating the electromagnetic field. Minor
difference has been found between considering only spectators and
all nucleons. For instance, for $b=10$~fm, protons in spectators
alone yield about 6\% smaller $B_{y}$ compared to protons in the
whole nucleus. For calculating the electromagnetic field in the rest
of this study, we average over 50,000 MC Glauber events for each impact
parameter setup to obtain a smooth geometric distribution of the source
charges.



\section{Spatial distributions of electromagnetic fields\label{sec:3}}

In this section we present our numerical results on the spatial distribution
of the electric and magnetic fields, compared between zero and finite
electric $(\sigma)$ and chiral magnetic ($\sigma_{\chi}$) conductivity
cases. Based on the previous discussions in Sec.~\ref{sec:2}, the
MC Glauber model is used to obtain the spacetime evolution profile
of electric charges 
for Au+Au collisions at $\sqrt{s_{\mathrm{NN}}}=200$~GeV with different
impact parameters, the Maxwell equations are solved for the electromagnetic
fields with finite $\sigma$ and $\sigma_{\chi}$, while the Lienard-Wiechert
solution is taken for the case of $\sigma=\sigma_{\chi}=0$.

\begin{figure*}
\begin{centering}
\includegraphics[width=7.7cm]{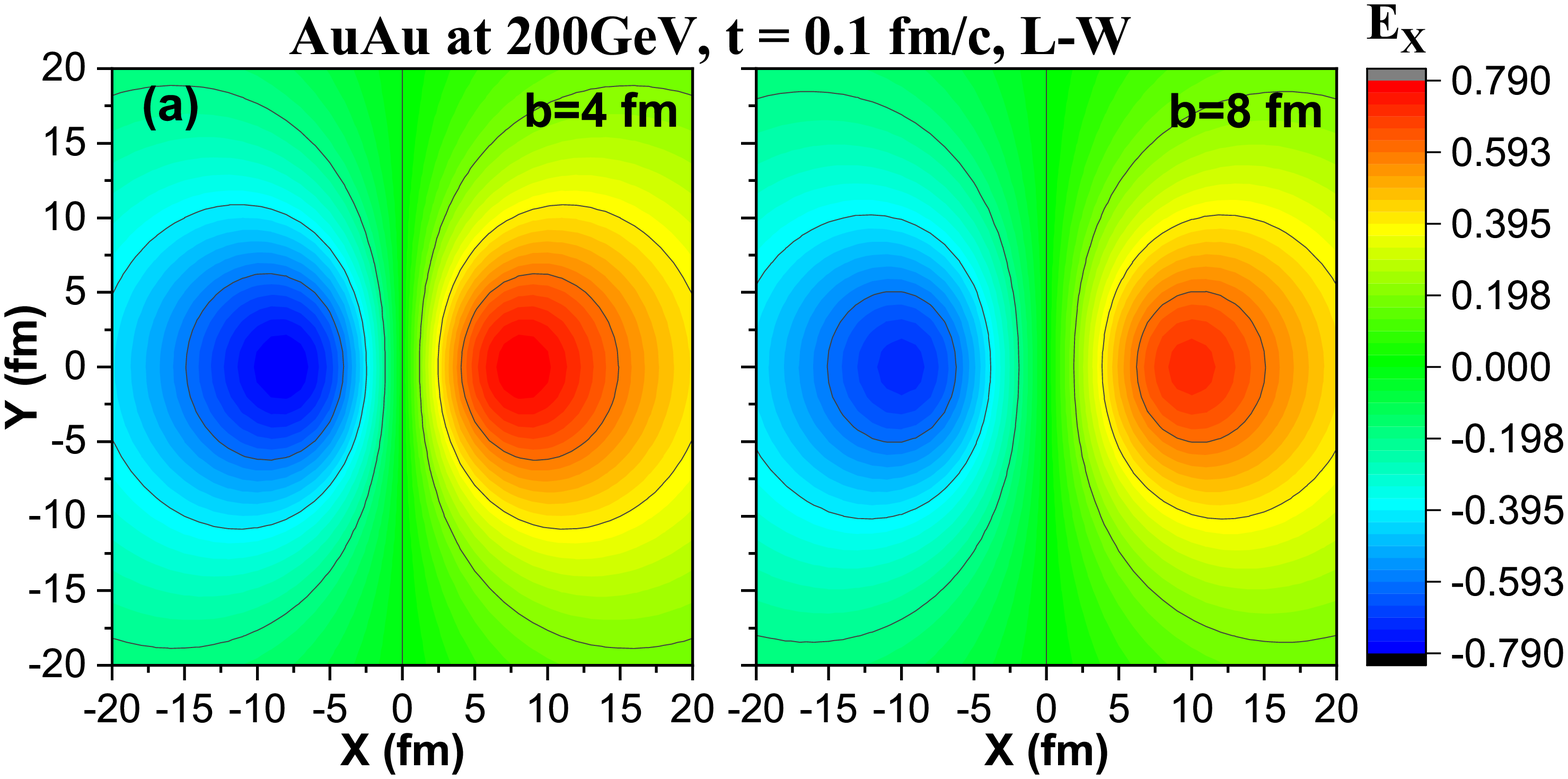} \includegraphics[width=7.7cm]{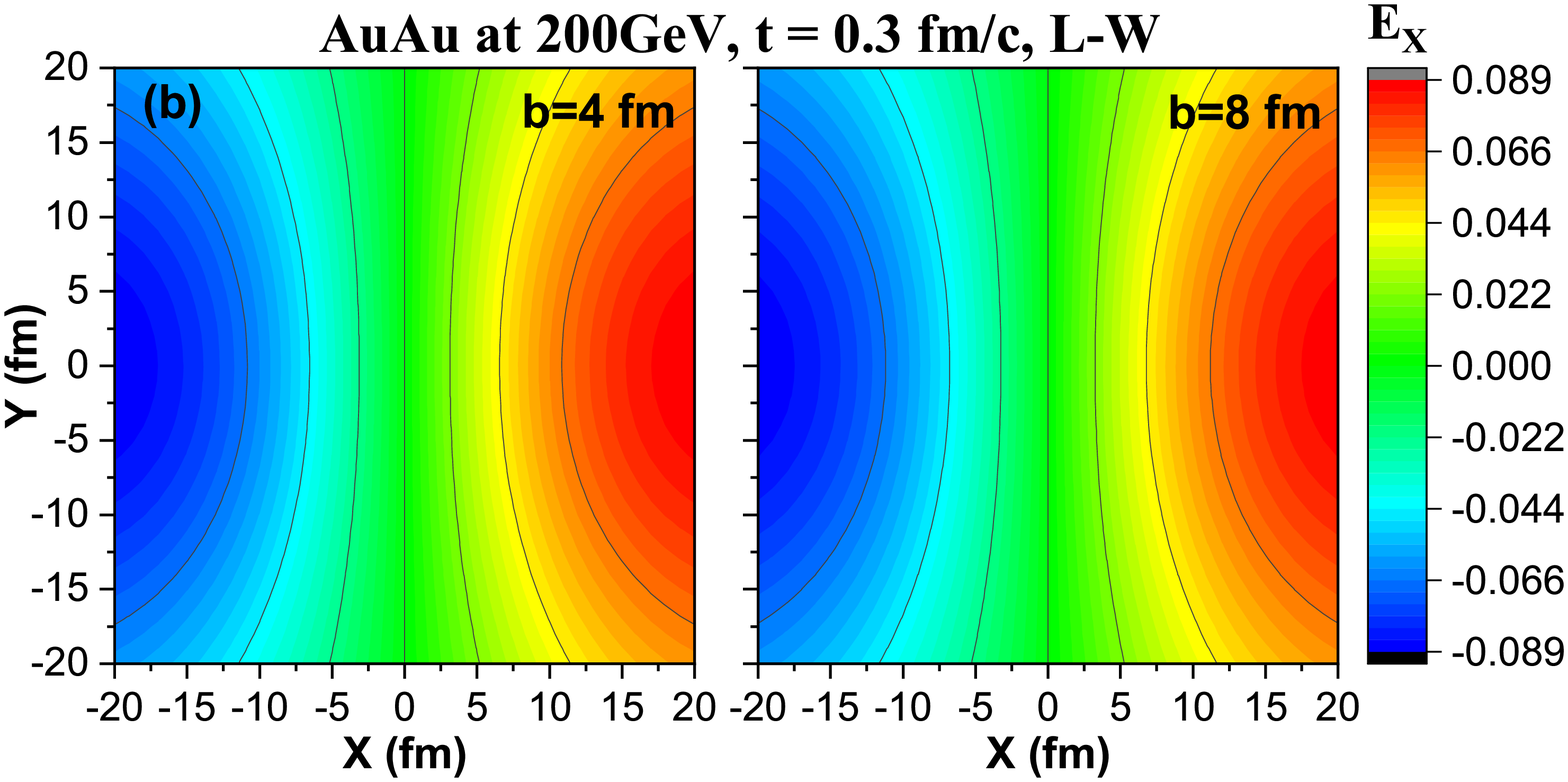} 
\par\end{centering}
\begin{centering}
\includegraphics[width=7.7cm]{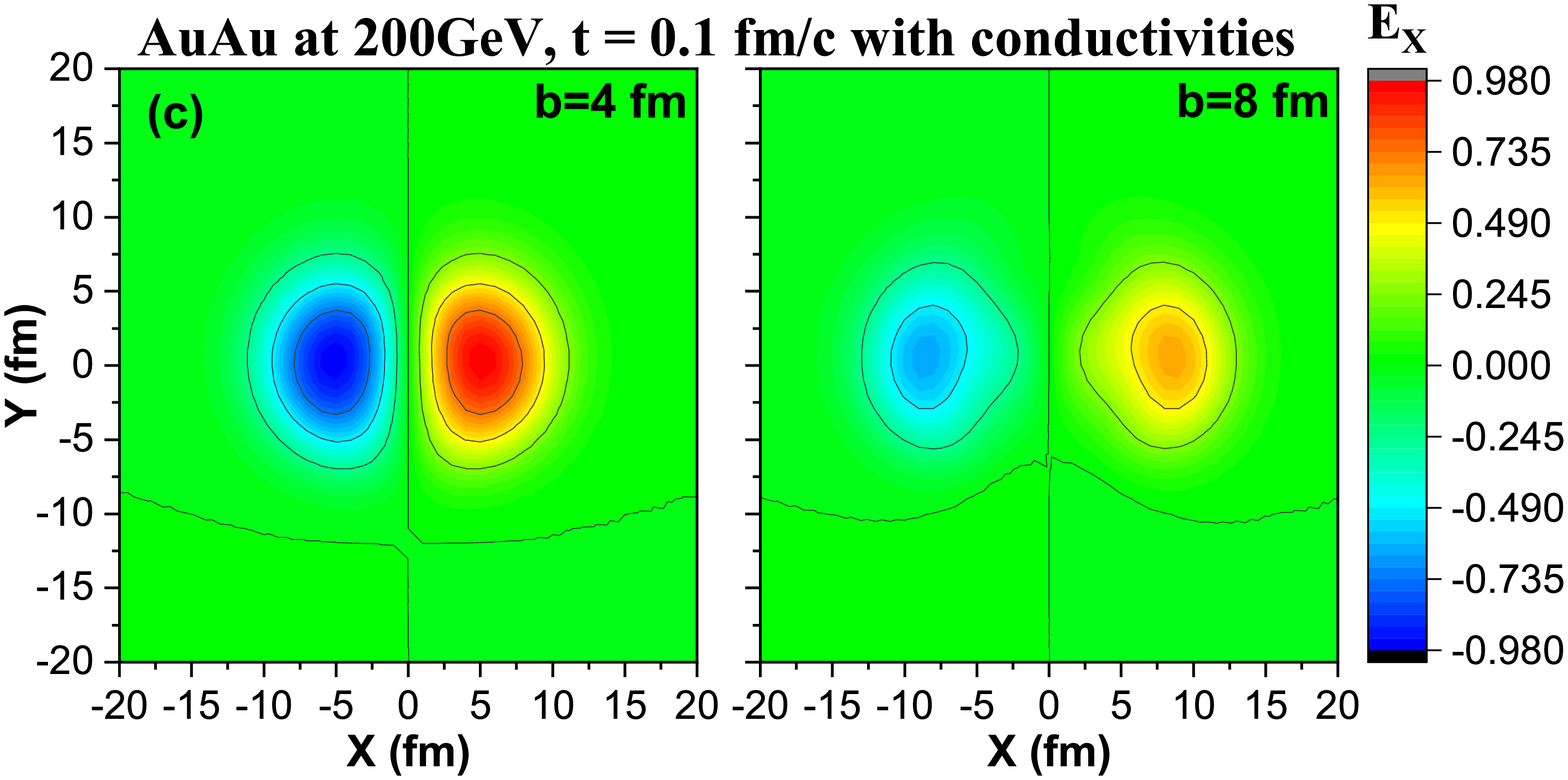} \includegraphics[width=7.7cm]{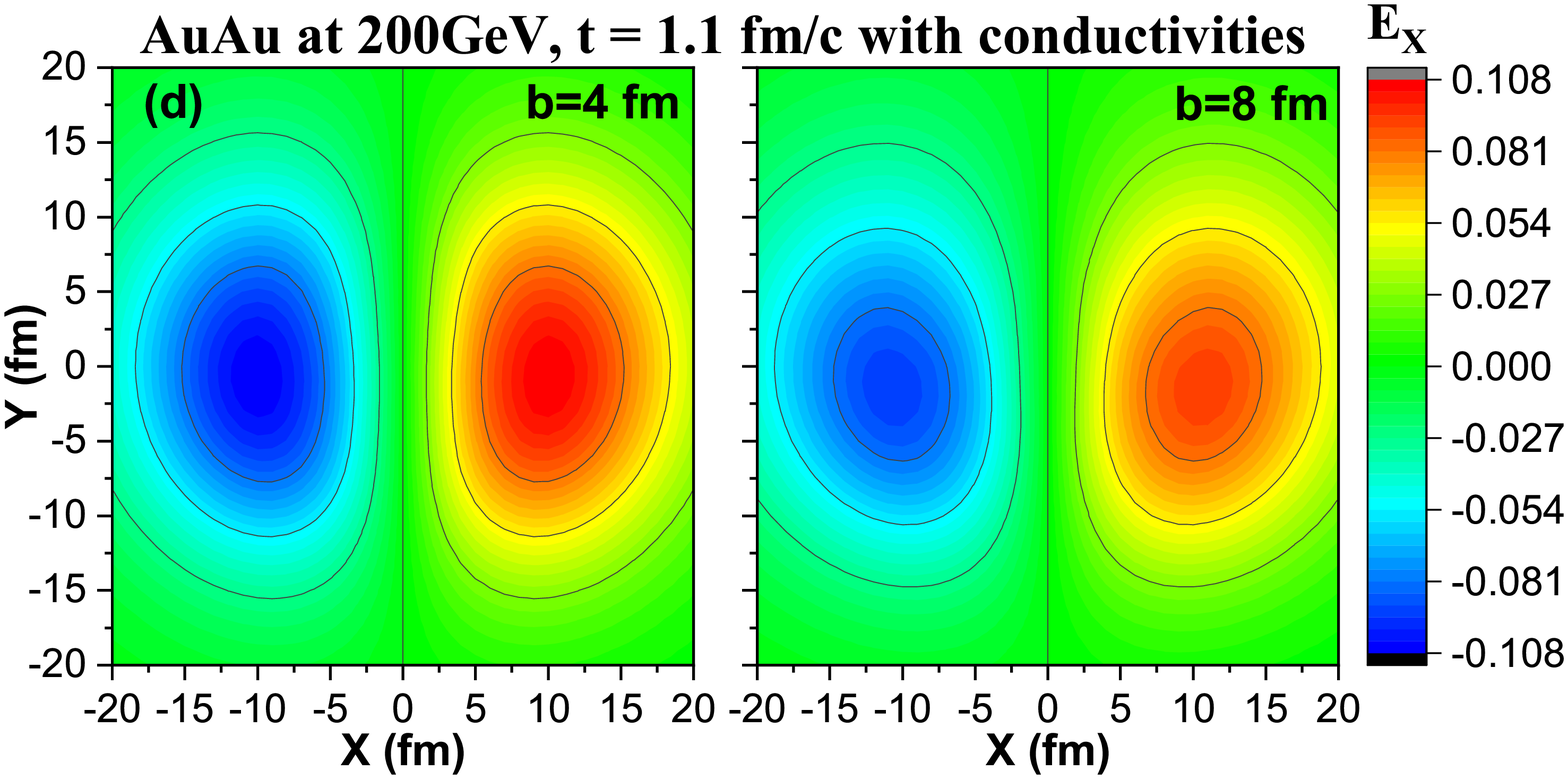} 
\par\end{centering}
\caption{\label{fig:Ex_spatial} (Color online) The spatial distributions of
$eE_{x}$ (in the unit of $m_{\pi}^{2}$) in 200~AGeV Au+Au collisions,
compared between zero \textit{vs.} finite conductivities, different
impact parameters and evolution times.}
\end{figure*}

\begin{figure*}
\begin{centering}
\includegraphics[width=7.7cm]{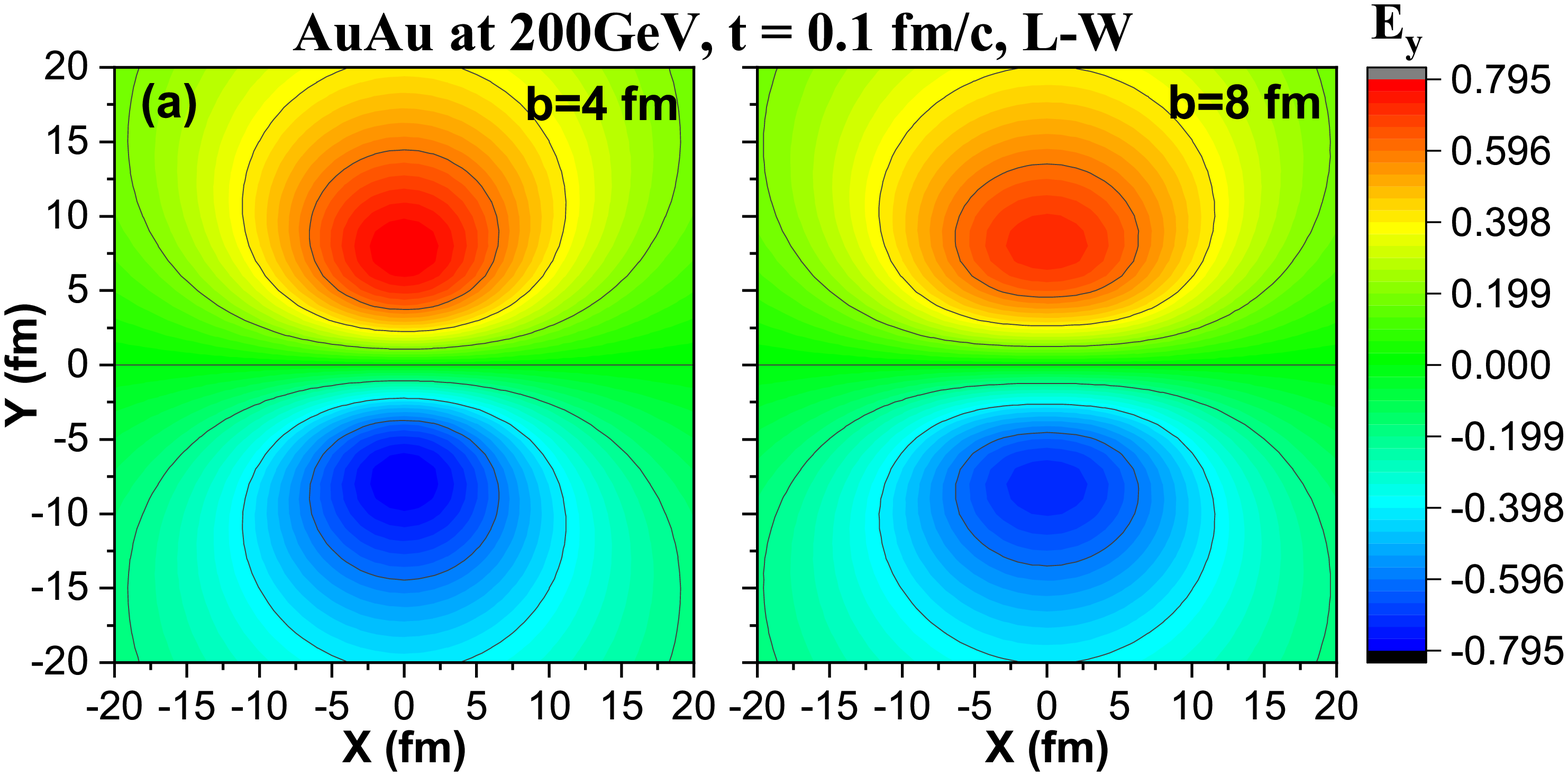} \includegraphics[width=7.7cm]{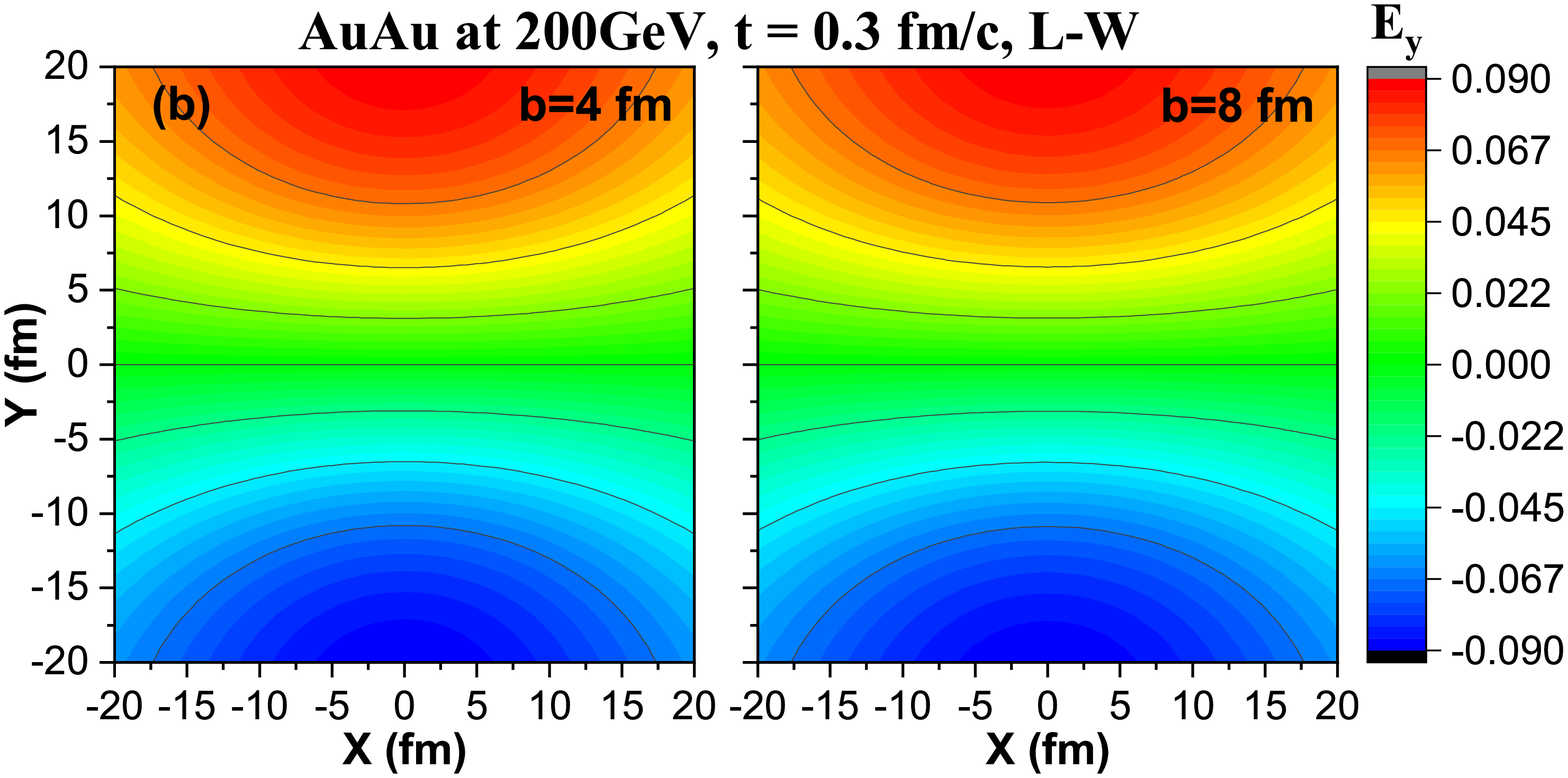} 
\par\end{centering}
\begin{centering}
\includegraphics[width=7.7cm]{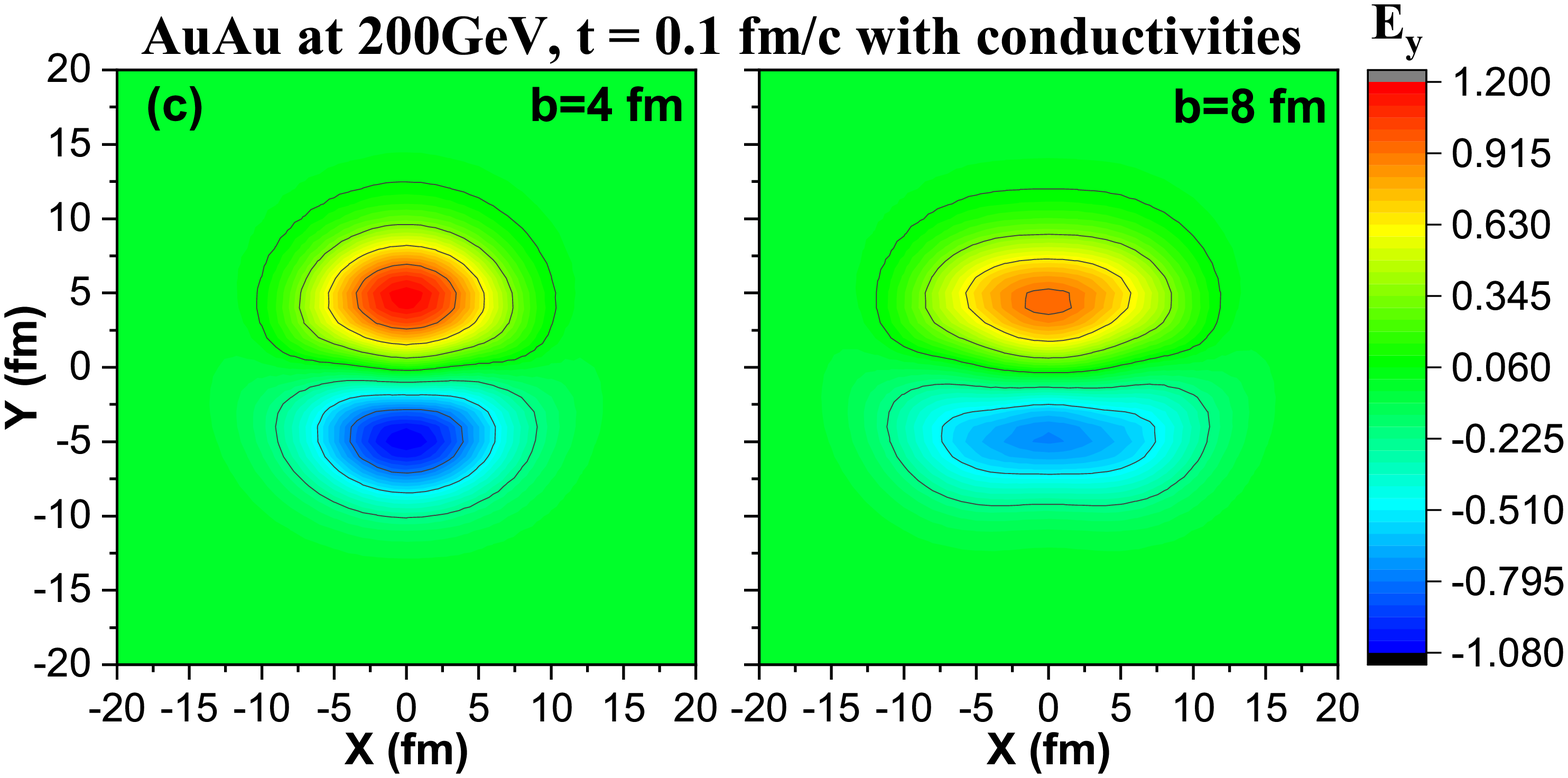} \includegraphics[width=7.7cm]{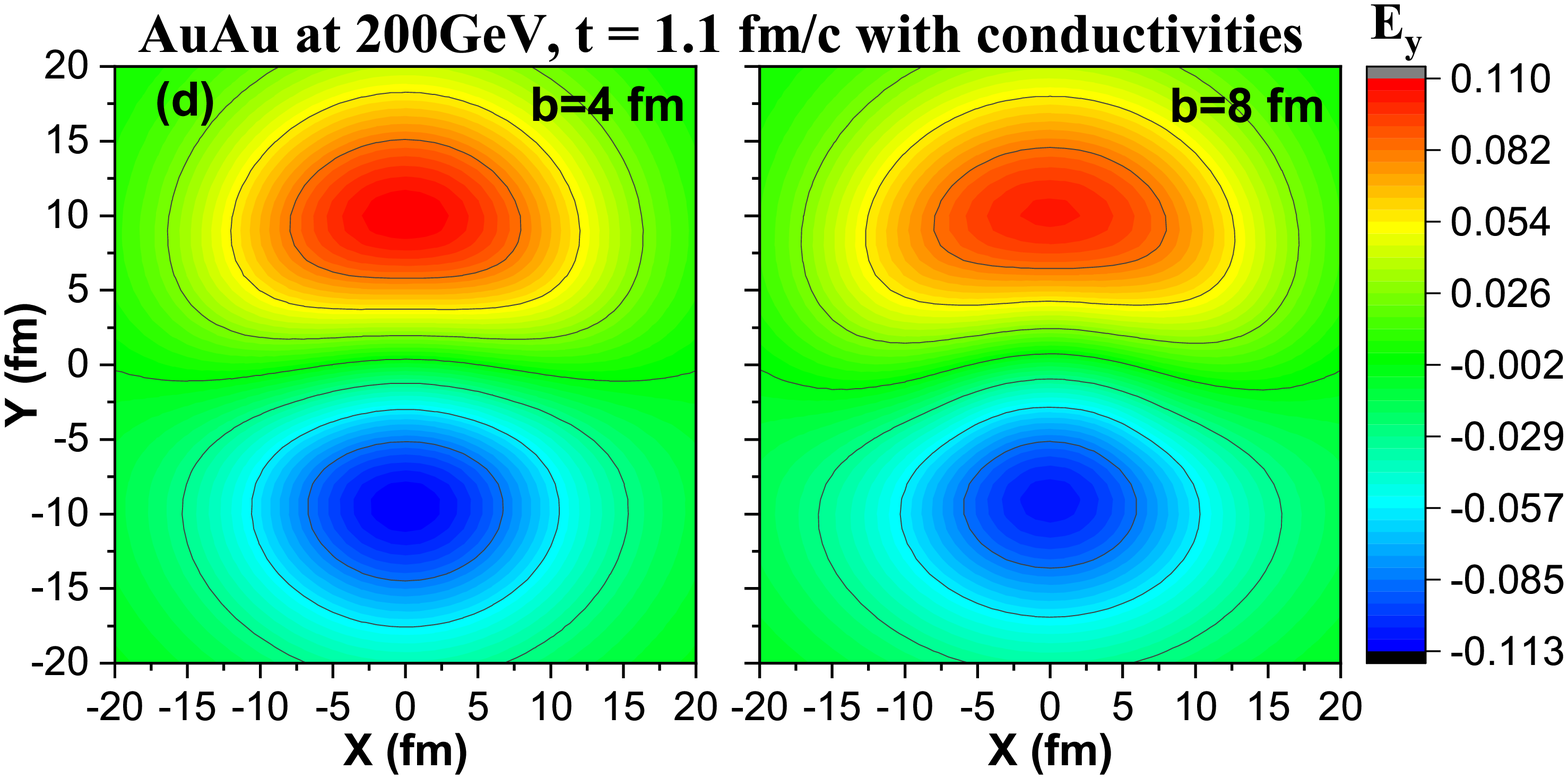} 
\par\end{centering}
\caption{\label{fig:Ey_spatial} (Color online) The spatial distributions of
$eE_{y}$ (in the unit of $m_{\pi}^{2}$) in 200~AGeV Au+Au collisions,
compared between zero \textit{vs.} finite conductivities, different
impact parameters and evolution times.}
\end{figure*}

\begin{figure*}
\begin{centering}
\includegraphics[width=7.7cm]{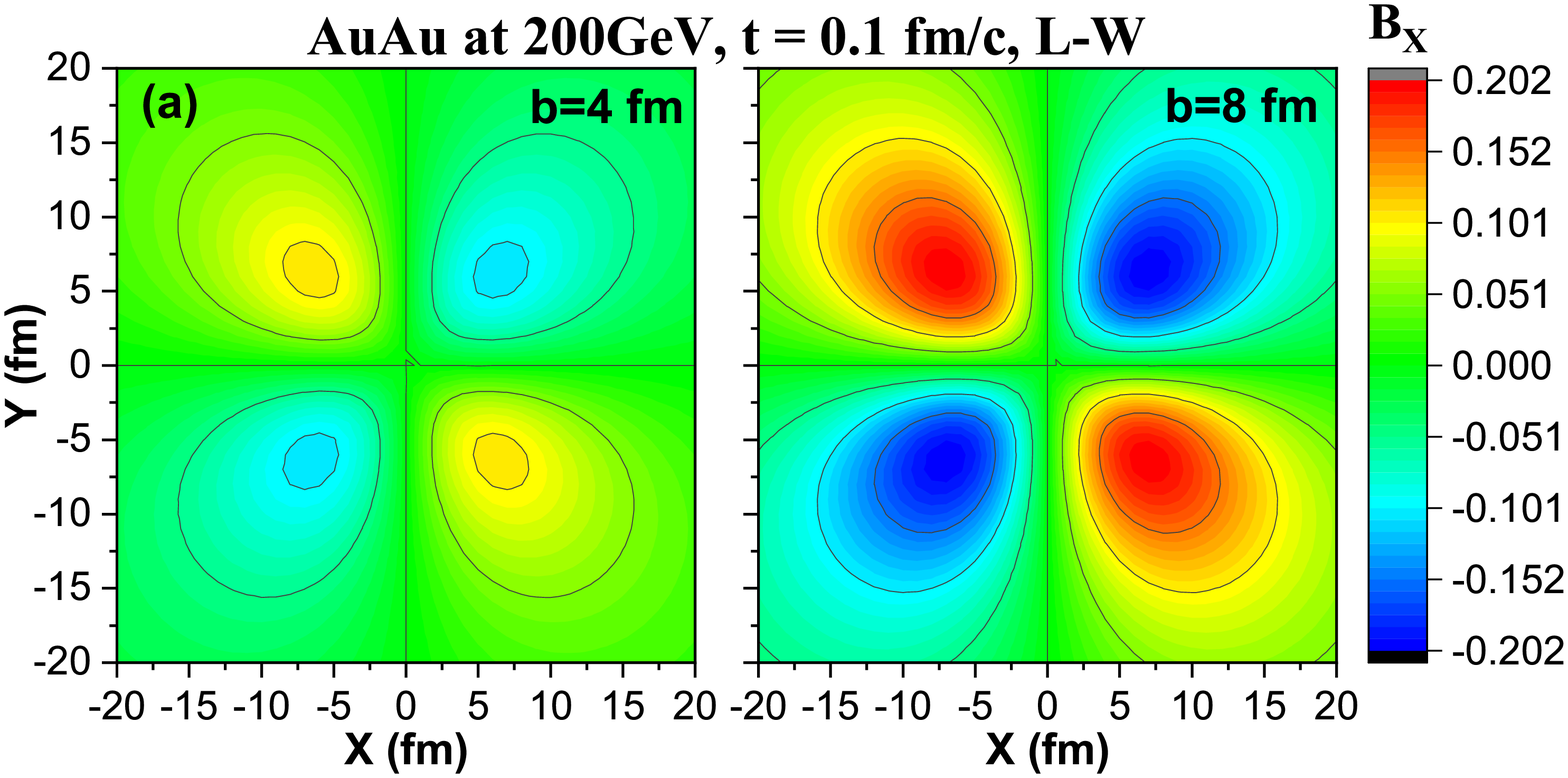} \includegraphics[width=7.7cm]{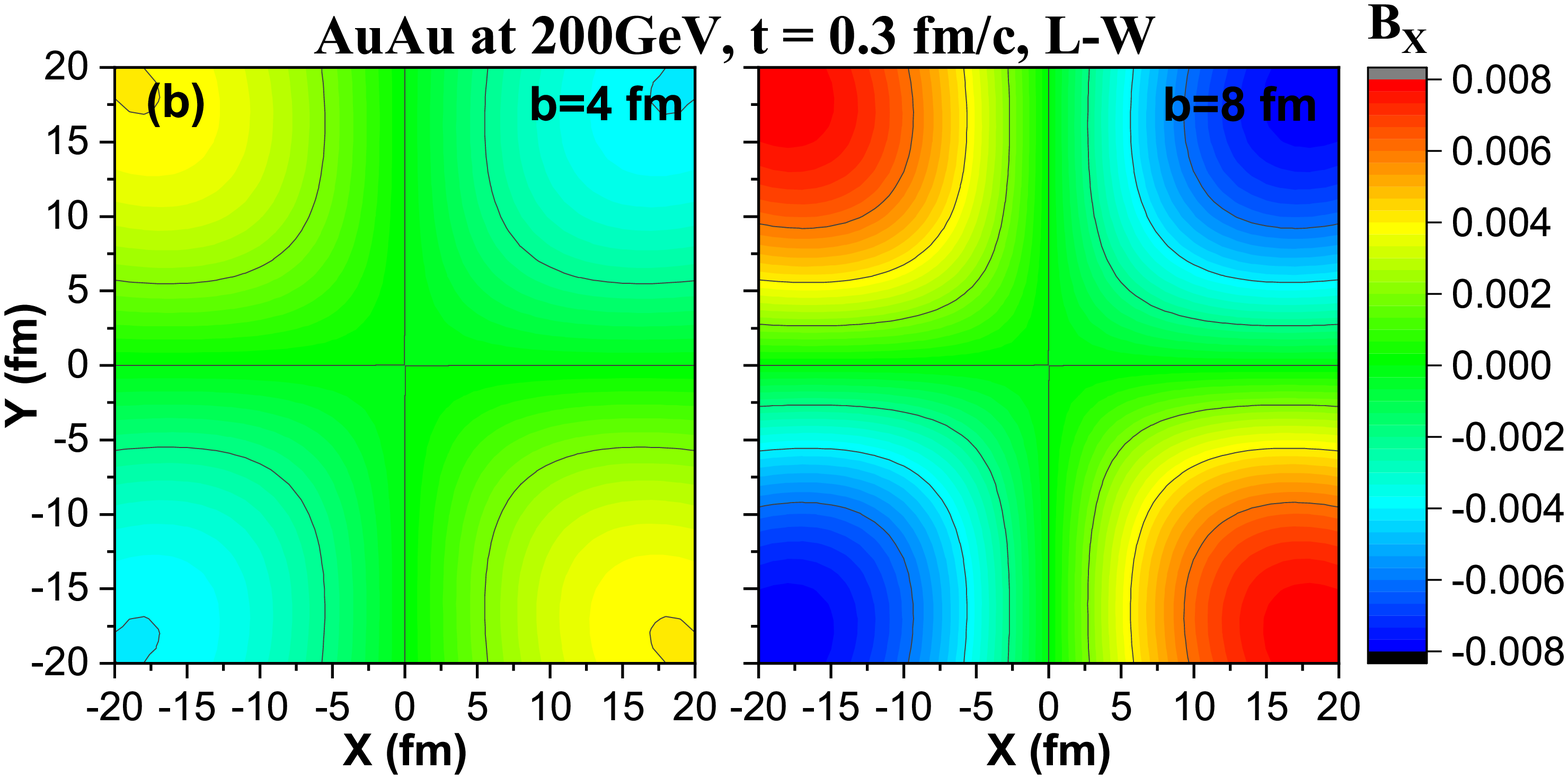} 
\par\end{centering}
\begin{centering}
\includegraphics[width=7.7cm]{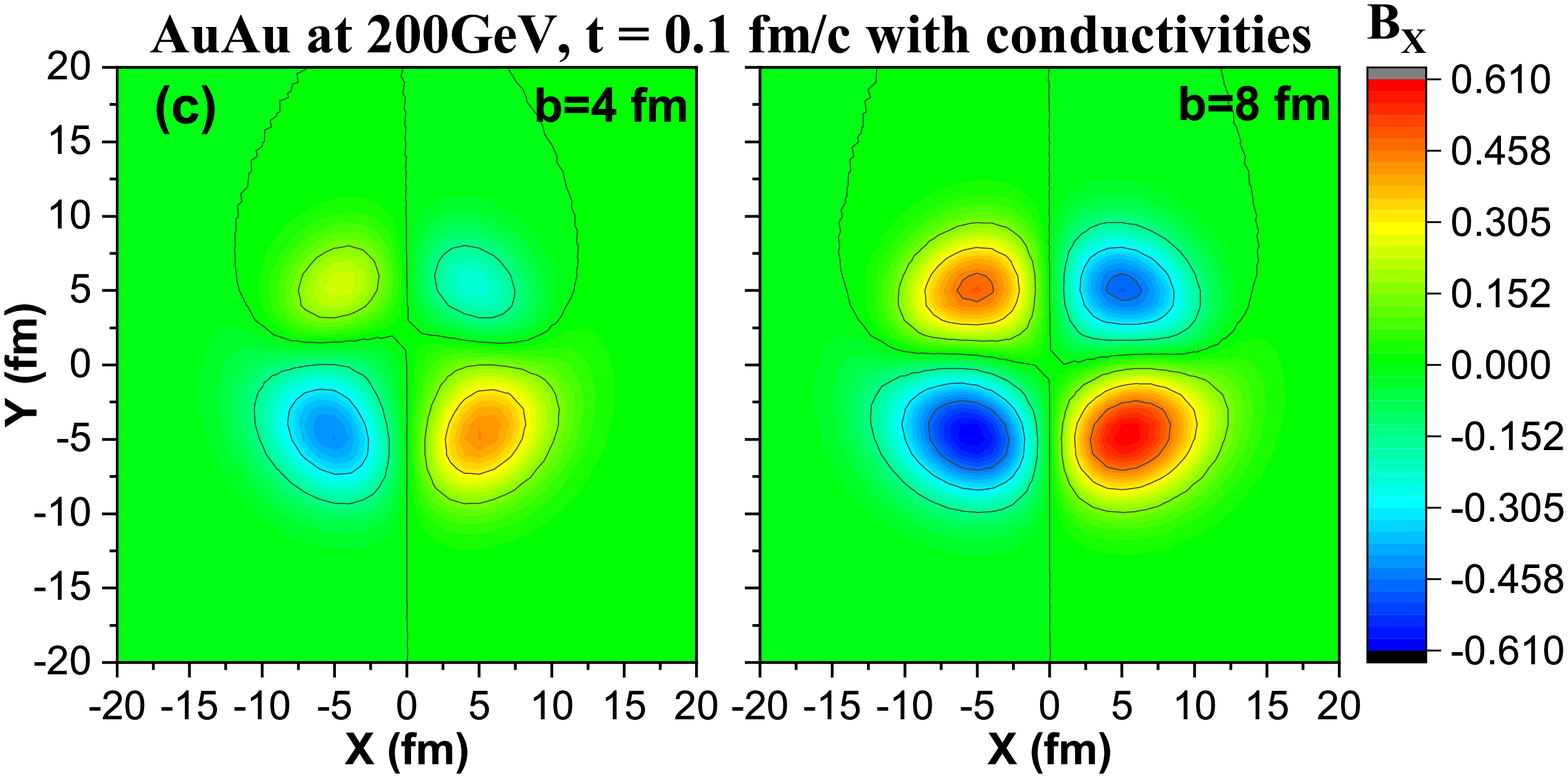} \includegraphics[width=7.7cm]{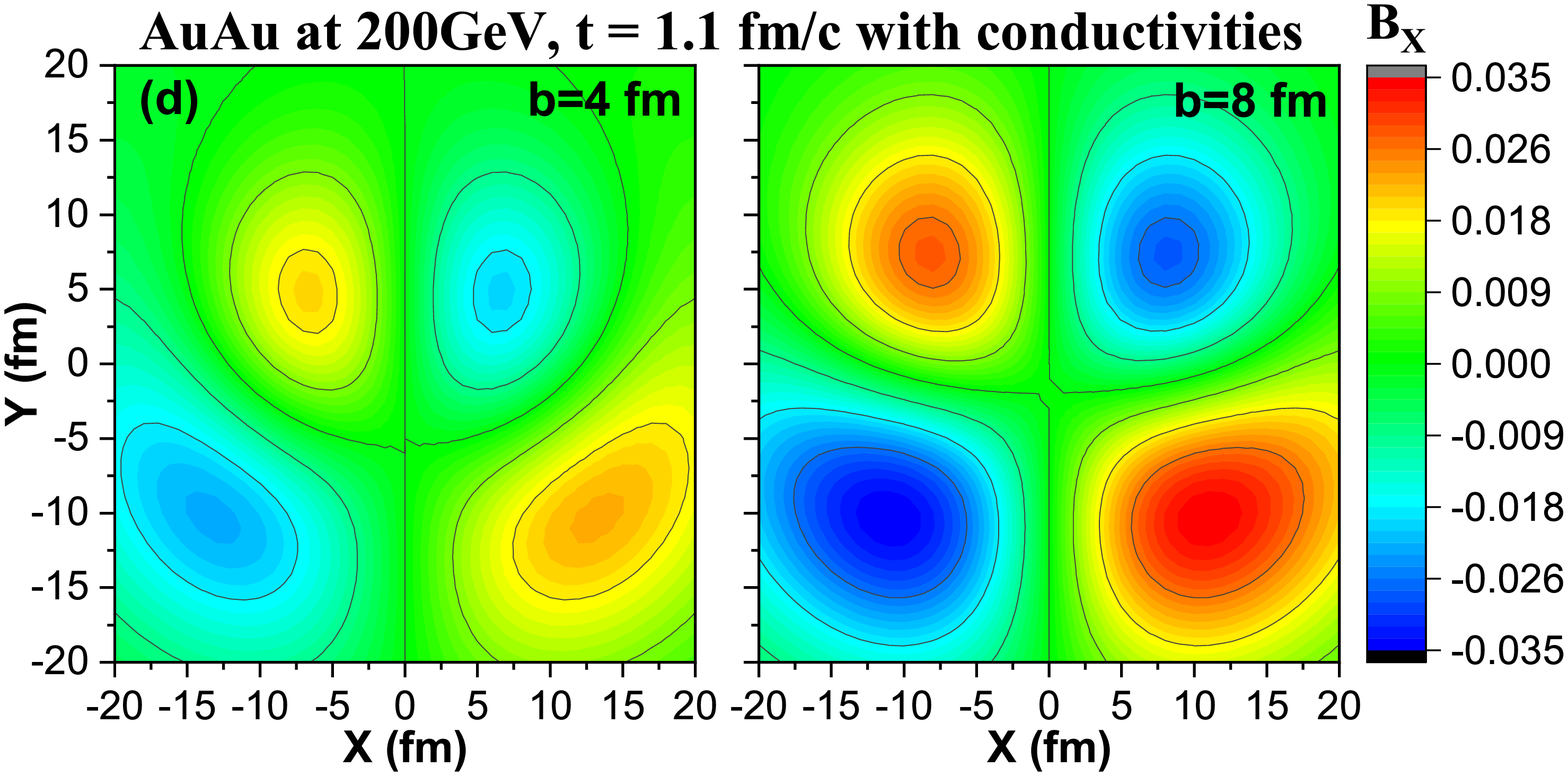} 
\par\end{centering}
\caption{\label{fig:Bx_spatial} (Color online) The spatial distributions of
$eB_{x}$ (in the unit of $m_{\pi}^{2}$) in 200~AGeV Au+Au collisions,
compared between zero \textit{vs.} finite conductivities, different
impact parameters and evolution times.}
\end{figure*}

\begin{figure*}
\begin{centering}
\includegraphics[width=7.7cm]{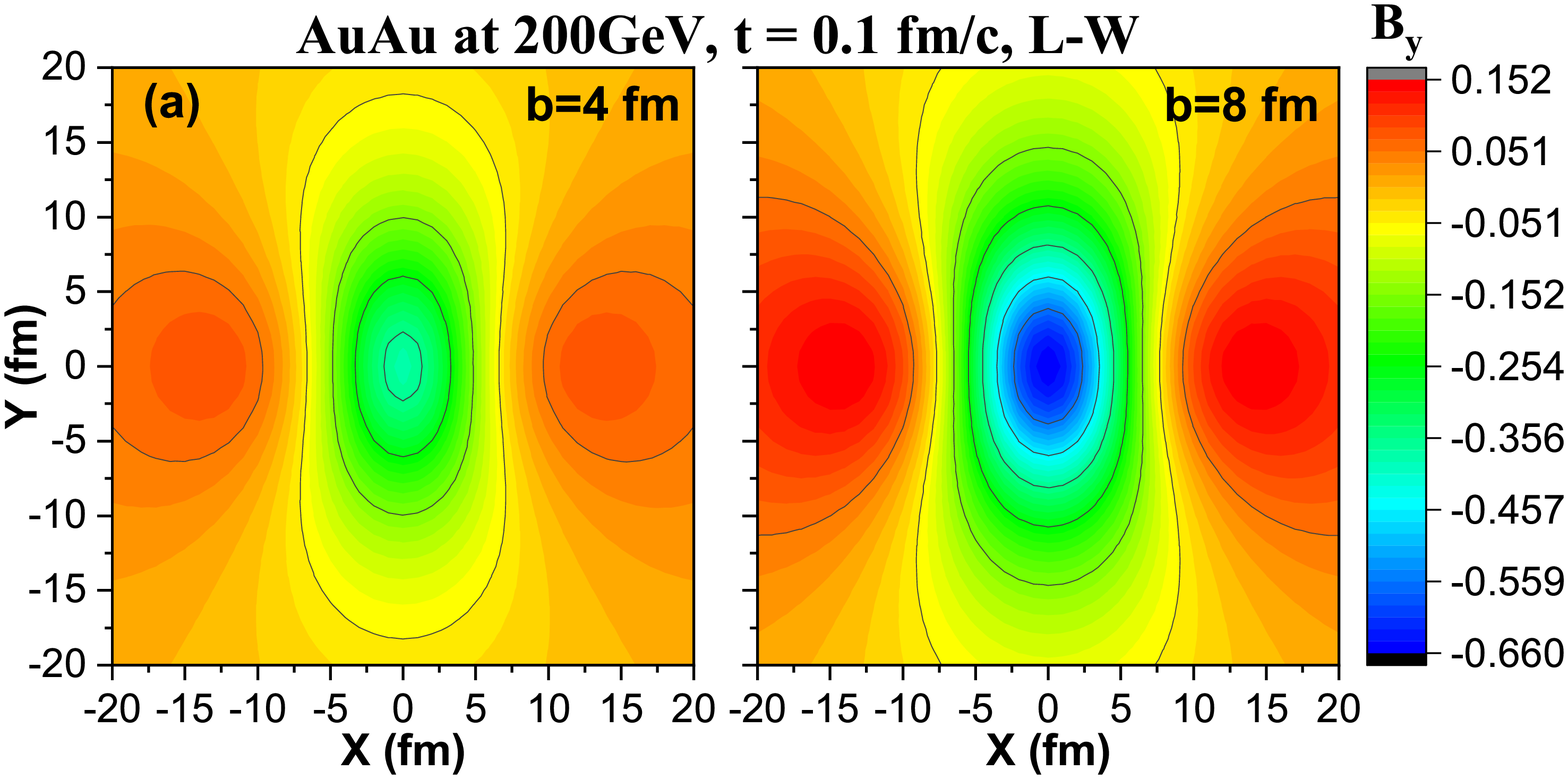} \includegraphics[width=7.7cm]{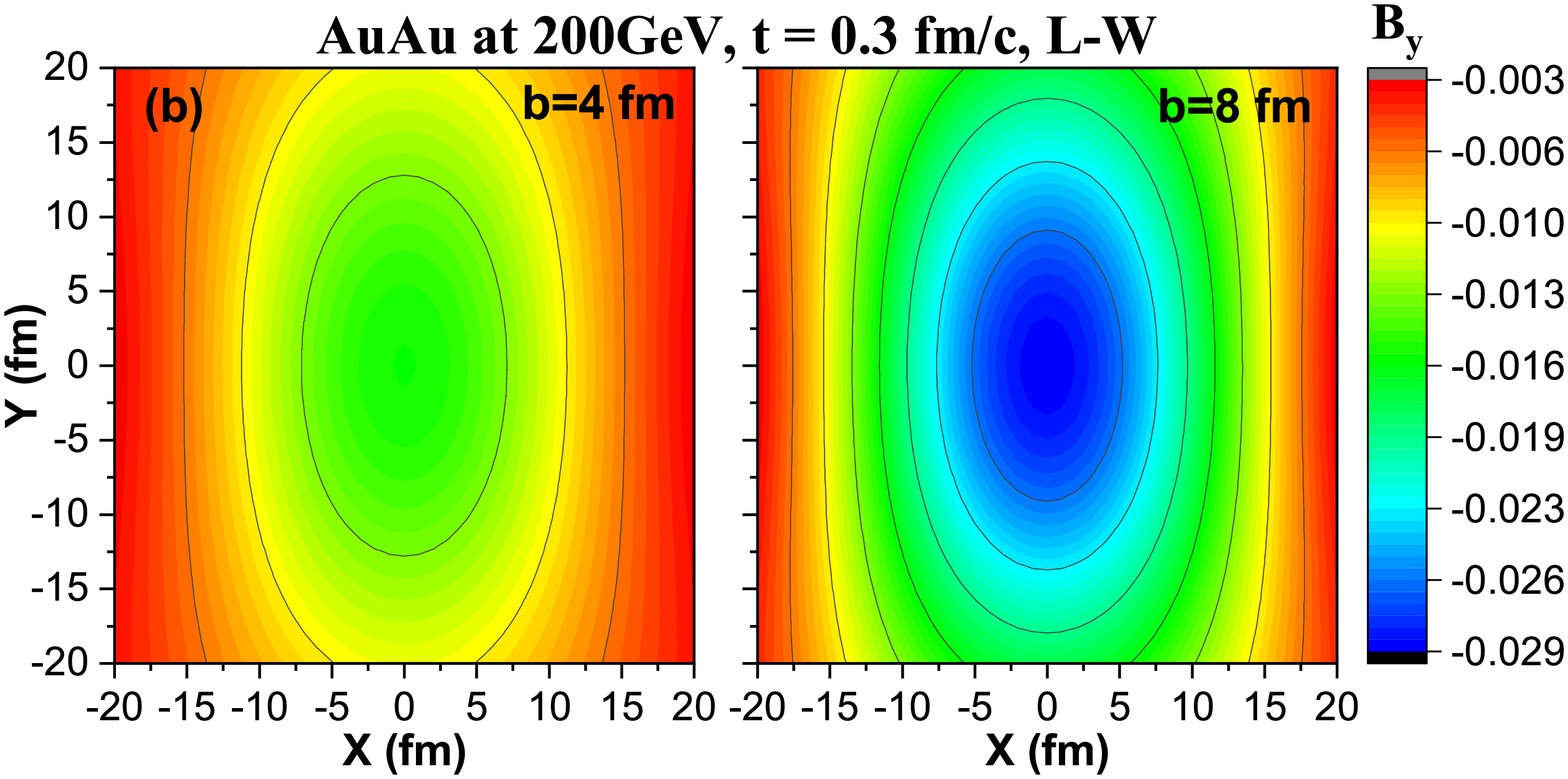} 
\par\end{centering}
\begin{centering}
\includegraphics[width=7.7cm]{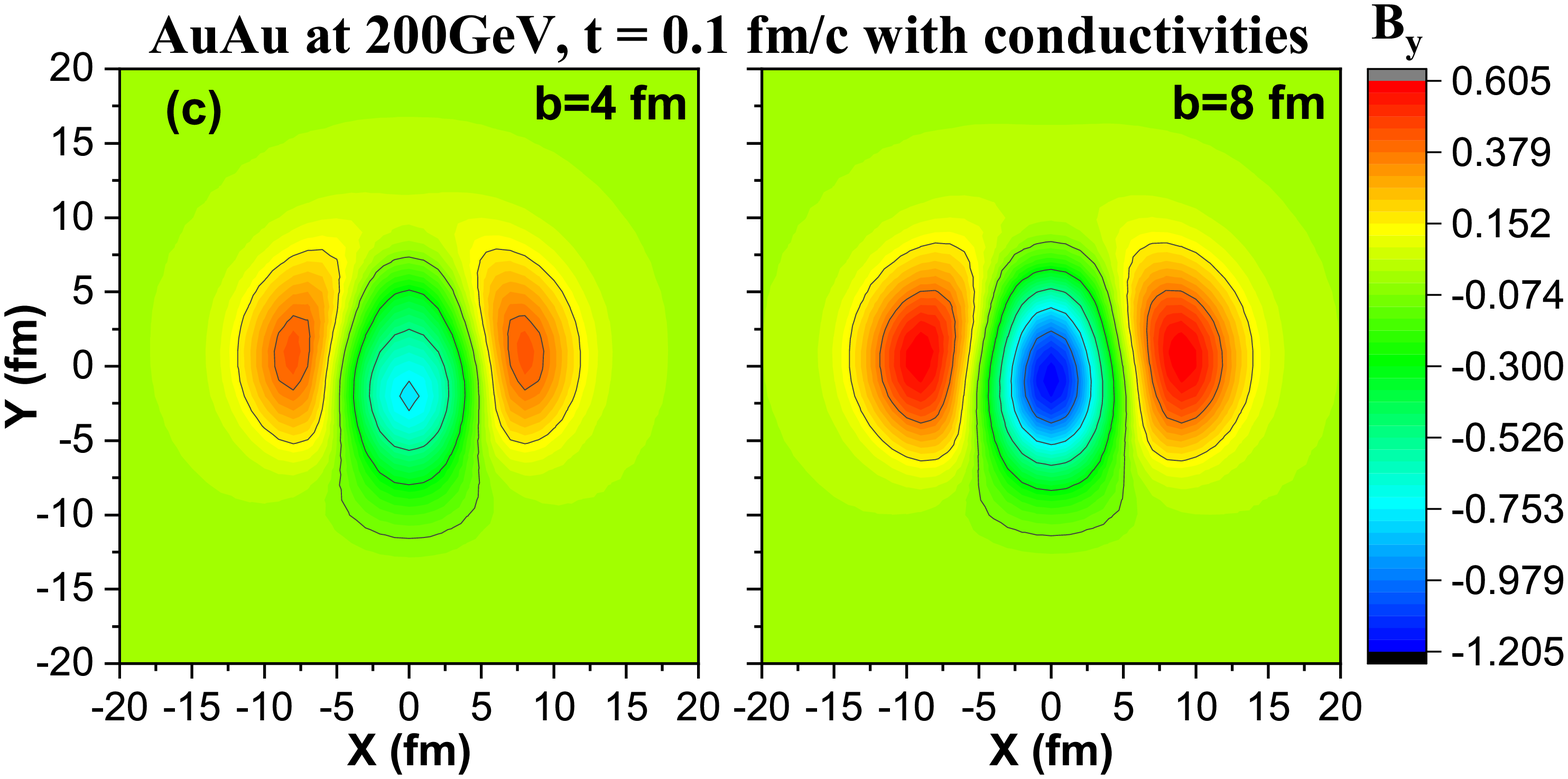} \includegraphics[width=7.7cm]{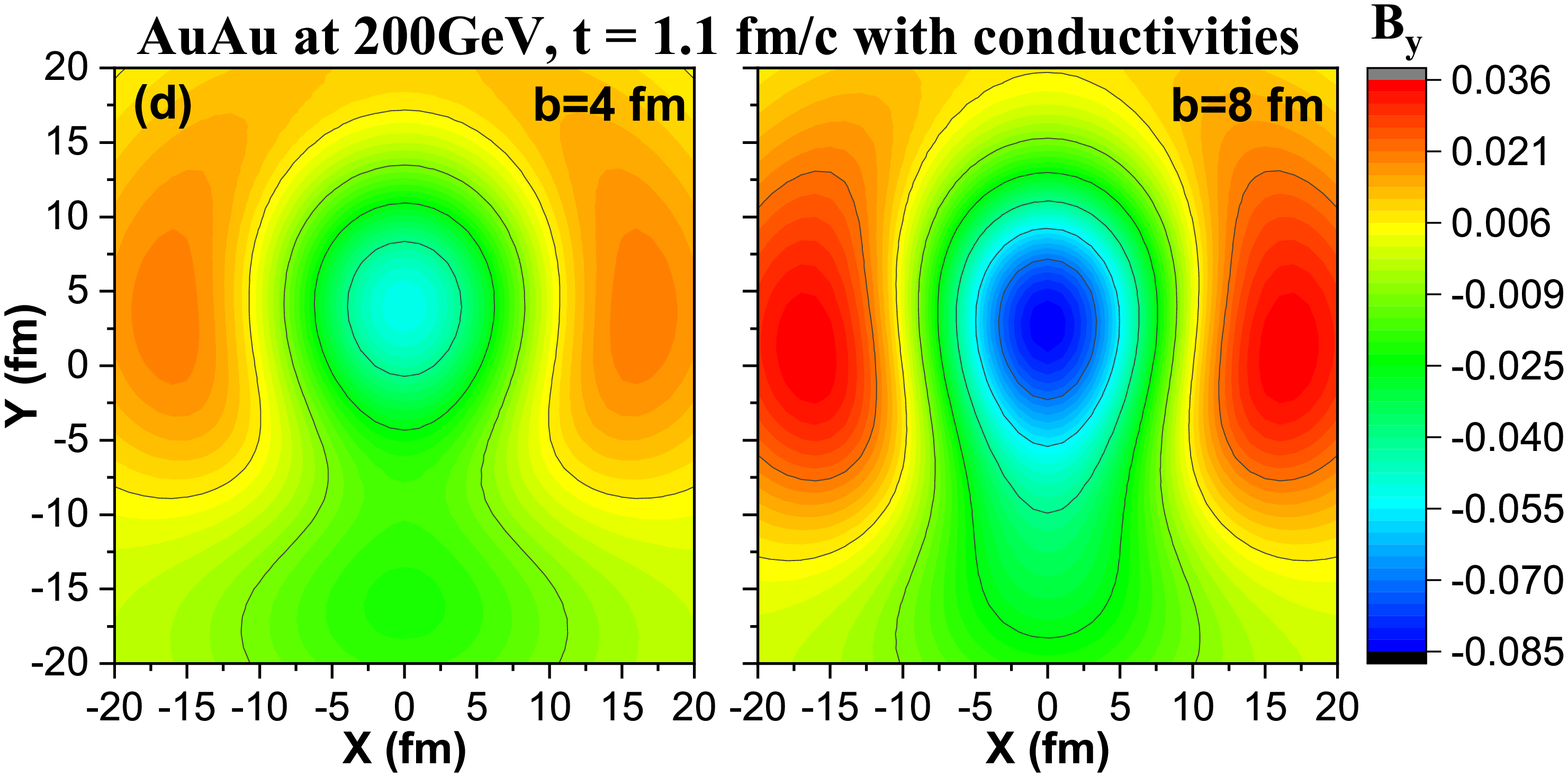} 
\par\end{centering}
\caption{\label{fig:By_spatial} (Color online) The spatial distributions of
$eB_{y}$ (in the unit of $m_{\pi}^{2}$) in 200~AGeV Au+Au collisions,
compared between zero \textit{vs.} finite conductivities, different
impact parameters and evolution times.}
\end{figure*}

Shown in Figs.~\ref{fig:Ex_spatial} - \ref{fig:By_spatial} are
the spatial distributions of $eE_{x}$, $eE_{y}$, $eB_{x}$, $eB_{y}$
respectively in Au+Au collisions at $\sqrt{s_{\mathrm{NN}}}=200$~GeV.
In each figure, we present the Lienard-Wiechert solution for vacuum
in the first row and results for finite conductivities in the second
row. Here the conductivities are taken as $\sigma=5.8$~MeV and $\sigma_{\chi}=1.5$~MeV
as used in Ref.~\cite{Li:2016tel}. Note that the value of $\sigma=5.8$~MeV is consistent with the lattice QCD result around the top temperature of the QGP at RHIC~\cite{Ding:2010ga,Aarts:2014nba}, which is expected to decrease together with the medium temperature as the QGP expands. In addition, there is no direct guidance of how to choose the value of $\sigma_{\chi}$ yet. Since the analytical solution of the electromagnetic field -- Eqs.~(\ref{eq:eq8}) and~(\ref{eq:eq9}) -- is obtained in the limit of $\sigma_\chi \ll \sigma$ in Ref.~\cite{Li:2016tel}, we take the value assigned in this original work. In the present study, we will only focus on investigating the effects of the electric and chiral magnetic conductivities with the given values above. Constraints on these values will be explored in a follow-up study where we connect electromagnetic effects to experimental observables.
In each row of the figures, two
snapshots of time evolution are presented. For the vacuum cases, due
to the rapid decay of the electromagnetic field, we present results
for $t=0.1$~fm/$c$ and 0.3~fm/$c$. On the other hand, $t=0.1$~fm/$c$
and 1.1~fm/$c$ are presented for the finite conductivity cases in
which the decay speed is much slower. And for each snapshot, results
for two impact parameters, $b=4$~fm and 8~fm are shown.

For the electric field presented in Figs.~\ref{fig:Ex_spatial} and~\ref{fig:Ey_spatial},
one observes its magnitude decreases as the impact parameter increases.
It is maximized at the most central collisions, as has been shown
in Ref.~\cite{Zhao:2019ybo}. As time evolves, the electric field
spreads out in space with a decreasing magnitude. Comparing between
the upper and lower panels, one can observe the electric field decays
much slower when the conductivities $\sigma$ and $\sigma_{\chi}$
are present. It is interesting to note that when $\sigma=\sigma_{\chi}=0$,
the spatial distributions of both $|E_{x}|$ and $|E_{y}|$ appear
symmetric with respect to both $x=0$ and $y=0$ axes. However, for
finite conductivities, these distributions are only symmetric about
the $x=0$ axis but asymmetric about $y=0$. This could be understood
with the non-zero azimuthal component $E_{\phi}$ with the presence
of $\sigma_{\chi}$. Similar to the illustration provided in Ref.~\cite{Li:2016tel},
if one assumes one proton travels in $-\hat{z}$ at $(-a,0,0)$ while
another travels in $+\hat{z}$ at $(a,0,0)$, the $E_{r}$ components
they generate according to Eq.~(\ref{eq:eq9}) will contribute to
the same sign of $E_{x}$ at two symmetric locations with respect
to the $y=0$ axis, e.g. $(b,c,0)$ and $(b,-c,0)$, while opposite
sign of $E_{y}$ at these two locations. To the contrary, the $E_{\phi}$
components from the two moving charges will generate opposite sign
of $E_{x}$ while same sign of $E_{y}$ at the two locations above.
As a result, the finite $E_{\phi}$ breaks the original symmetry of
$E_{x}(y)=E_{x}(-y)$ and $E_{y}(y)=-E_{y}(-y)$ at zero conductivities.


Some similar features can be observed in Figs.~\ref{fig:Bx_spatial}
and~\ref{fig:By_spatial} for the spatial distributions of the magnetic
field, such as the slower decay of $|B_{x}|$ and $|B_{y}|$ and their
spread into space after finite $\sigma$ and $\sigma_{\chi}$ are
included, as well as their broken symmetry with respect to the reaction
plane when finite conductivities are present. However, different from
the electric field, it is the radial component $B_{r}$ in Eq.~(\ref{eq:eq8}),
determined by $\sigma_{\chi}$, that breaks the original $B_{x}(y)=-B_{x}(-y)$
and $B_{y}(y)=B_{y}(-y)$ symmetry at zero conductivities. In addition,
different patterns of the spatial distribution can also be found between
electric and magnetic fields. And opposite to the electric field,
the magnetic field increases as the impact parameter increases.

\begin{figure*}
\begin{centering}
\includegraphics[width=7.7cm]{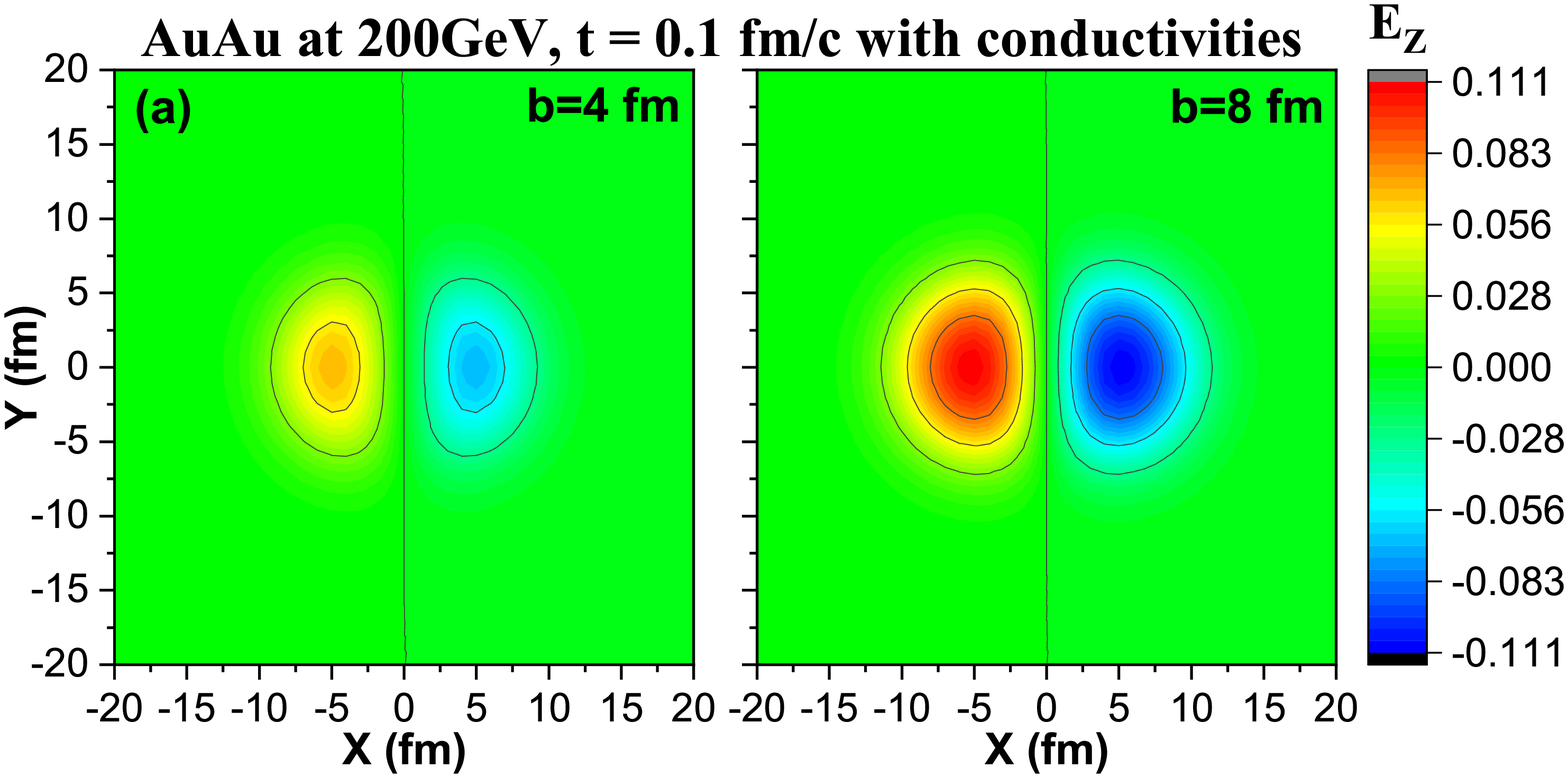} \includegraphics[width=7.7cm]{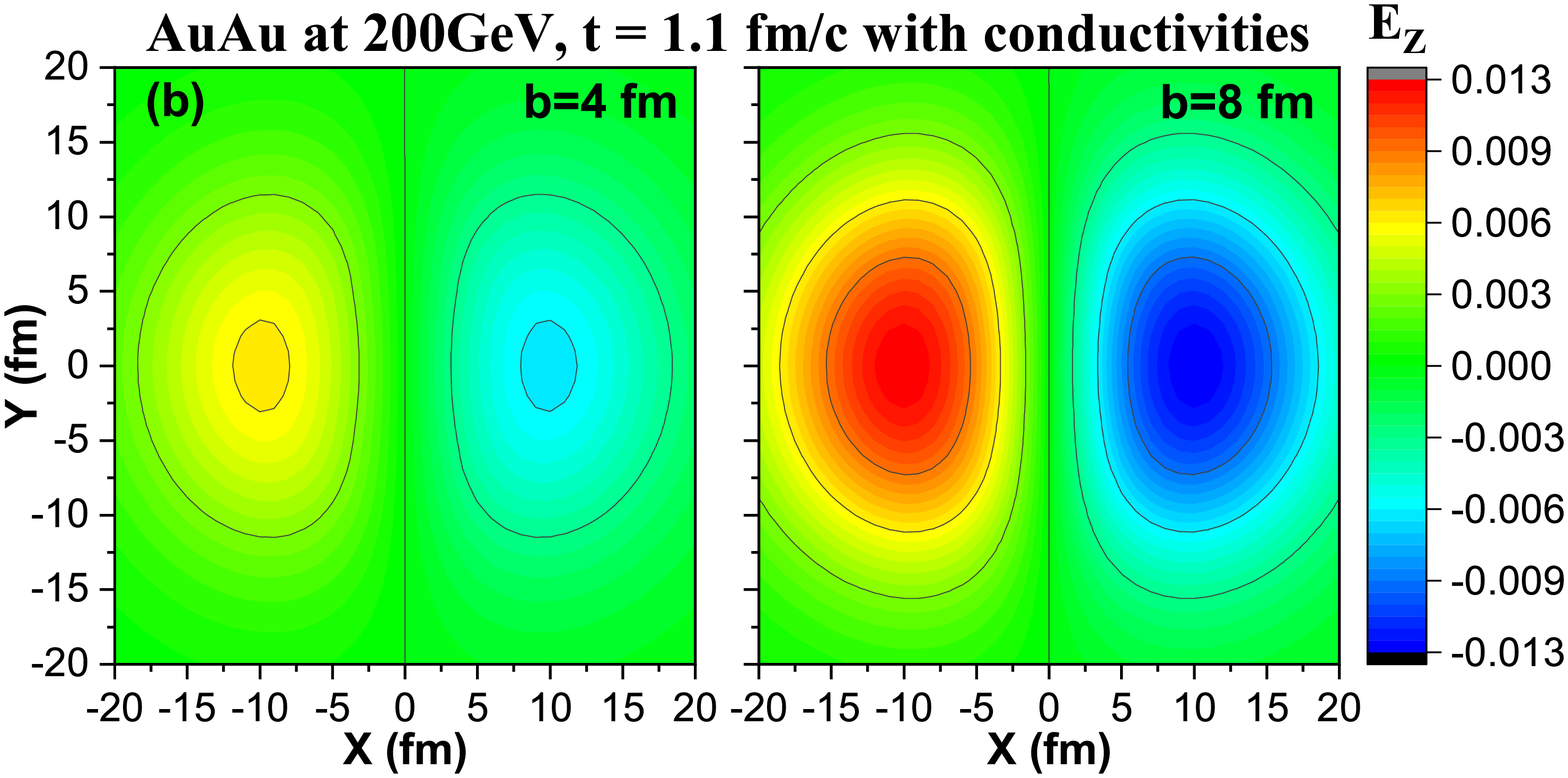} 
\par\end{centering}
\begin{centering}
\includegraphics[width=7.7cm]{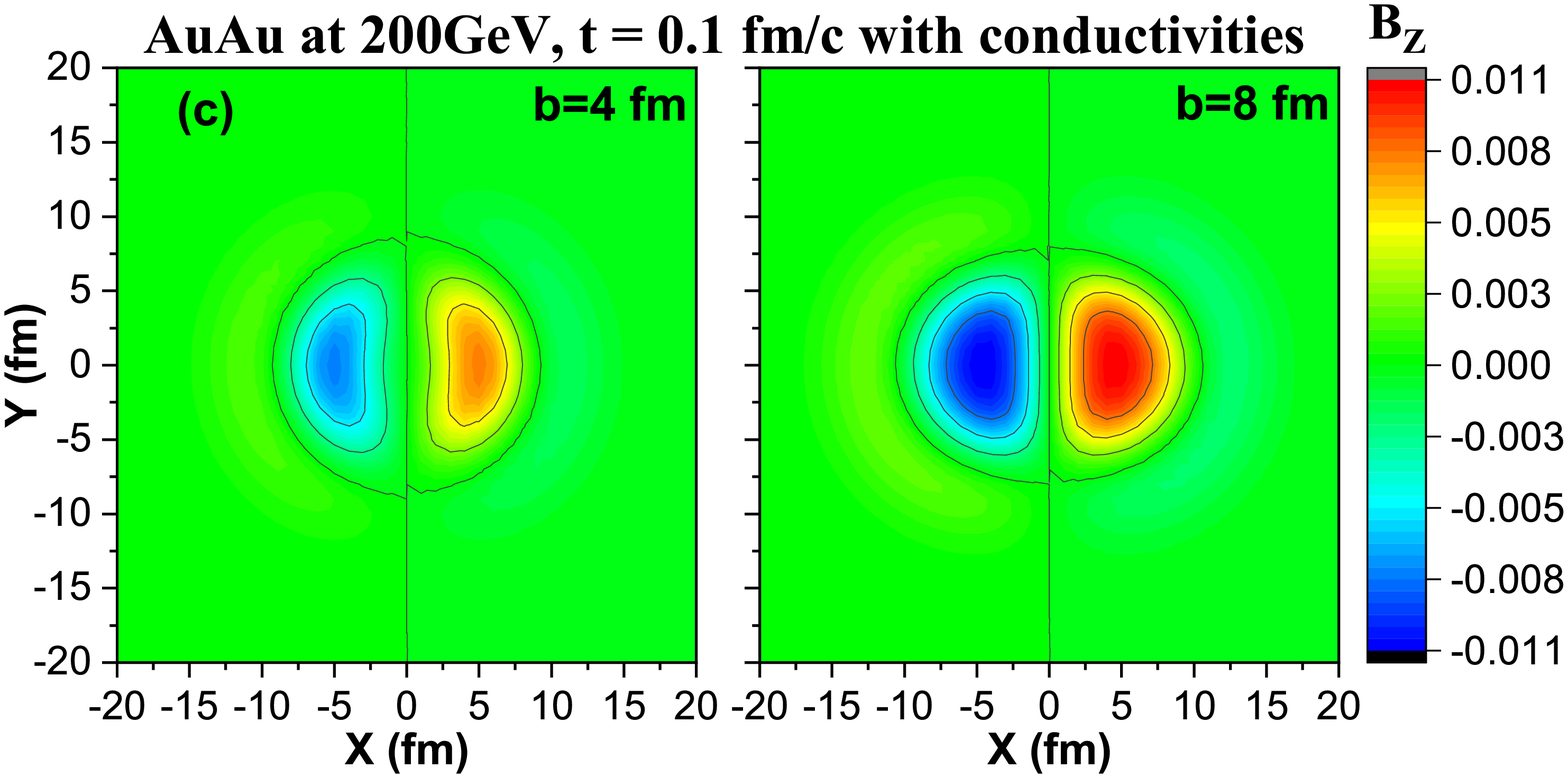} \includegraphics[width=7.7cm]{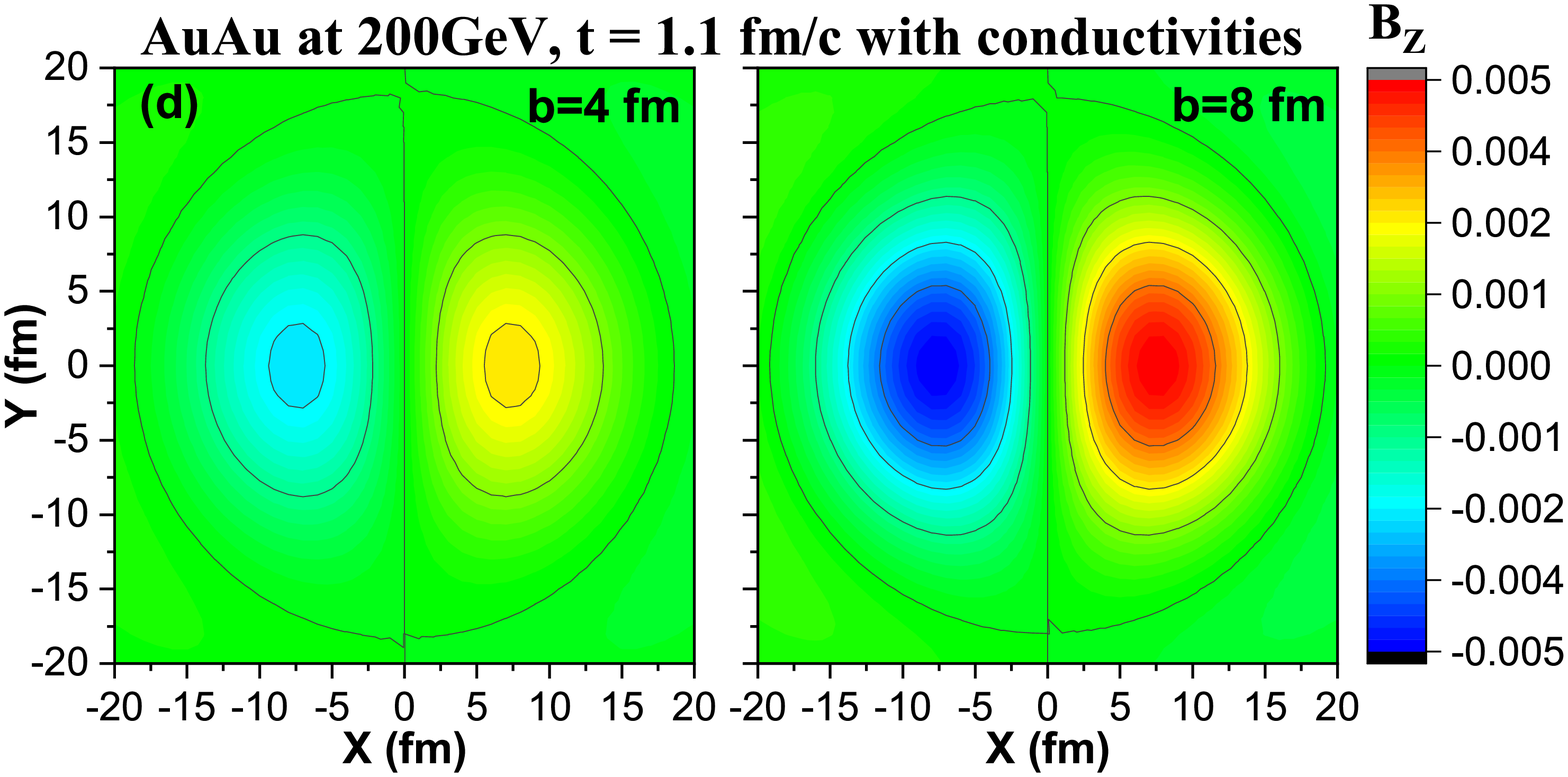} 
\par\end{centering}
\caption{\label{fig:EzBz_spatial} (Color online) The spatial distributions
of $eE_{z}$ (upper panels) and $eB_{z}$ (lower panels) (in the unit
of $m_{\pi}^{2}$) in 200~AGeV Au+Au collisions in the presence of
finite conductivities, compared between different impact parameters
and evolution times.}
\end{figure*}

In Fig.~\ref{fig:EzBz_spatial}, we present the spatial distributions
of the electric (upper panels) and magnetic (lower panels) field in
the longitudinal direction. Unlike the transverse components, even
in the presence of the electric and chiral magnetic conductivities,
$E_{z}$ and $B_{z}$ distributions still appear symmetric about both
$x=0$ and $y=0$ axes. This could be understood with the $B_z$ and $E_z$ components directly given by Eqs.~(\ref{eq:eq8}) and~(\ref{eq:eq9}). Two protons moving along $\pm\hat{z}$ at $(\pm a, 0,0)$ yield $F_{z}(y)=F_{z}(-y)$ and $F_{z}(x)=-F_{z}(-x)$, with $F_z$ for both $B_z$ and $E_z$.
And compared to their corresponding transverse components, the magnitudes
of electric and magnetic fields are much smaller in the longitudinal
direction, as has been suggested in Refs.~\cite{Bzdak:2011yy,Deng:2012pc,Zhao:2017rpf}. 

\begin{figure*}
\begin{centering}
\includegraphics[width=7.7cm]{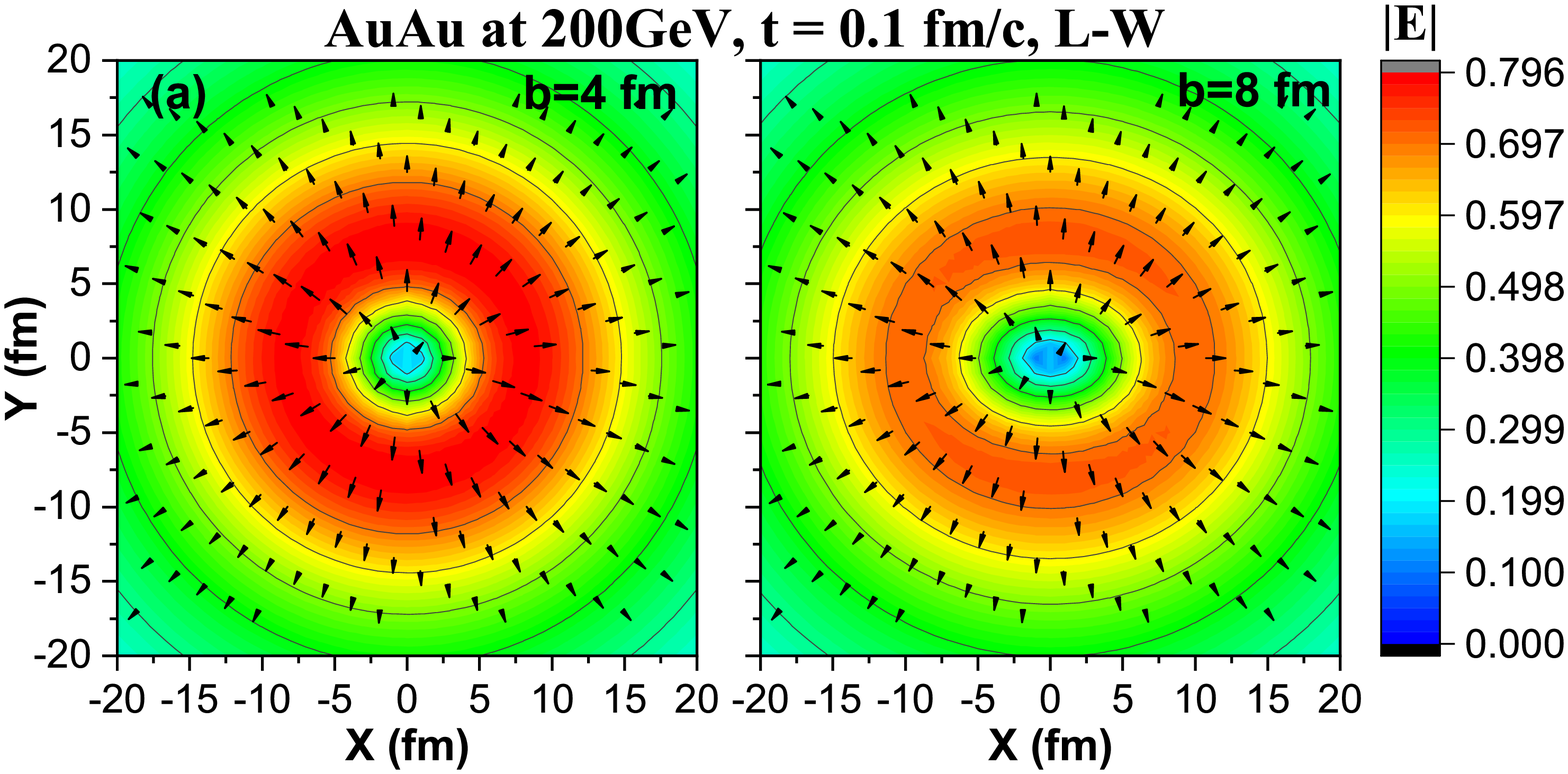} \includegraphics[width=7.7cm]{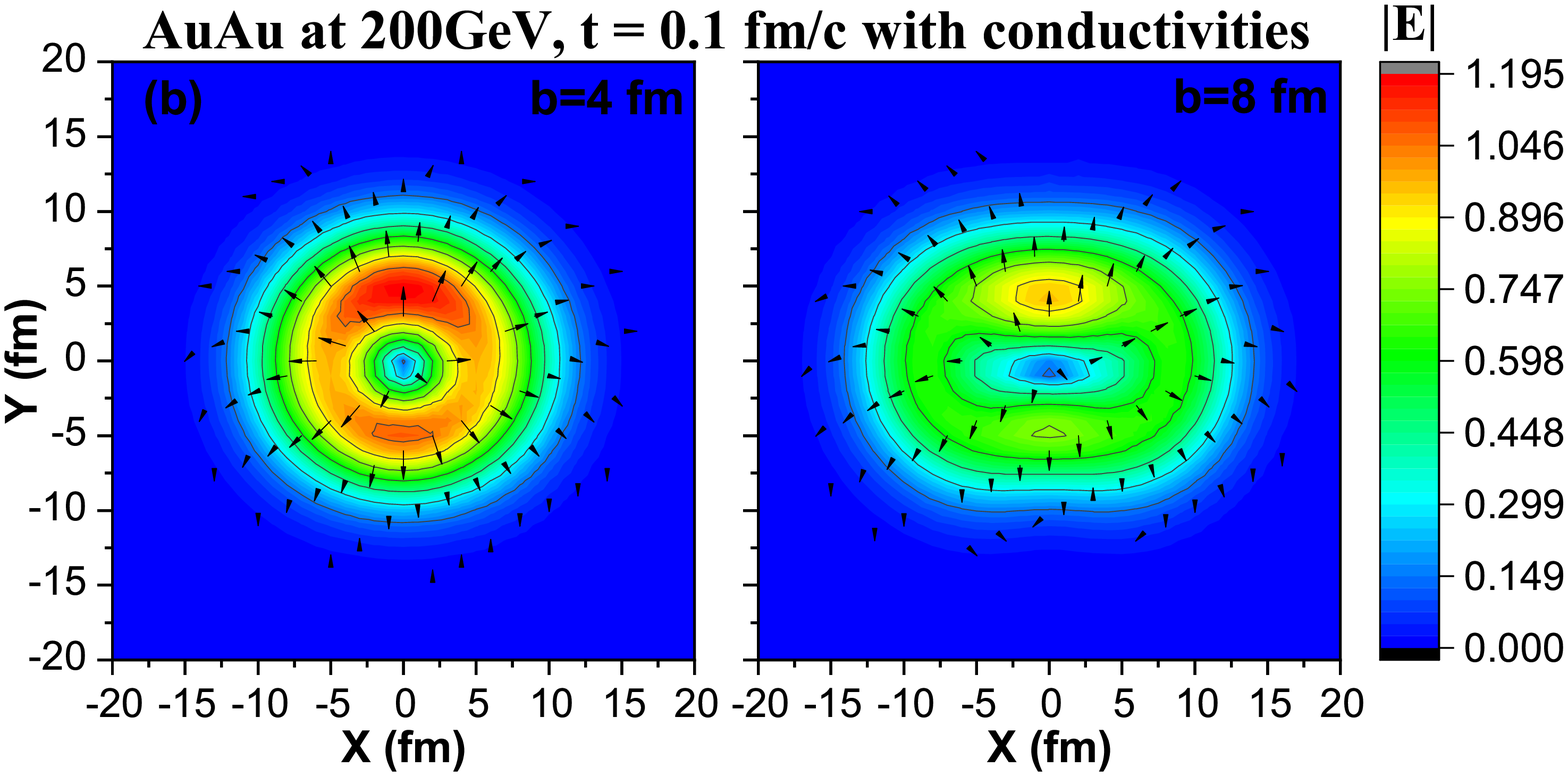} 
\par\end{centering}
\begin{centering}
\includegraphics[width=7.7cm]{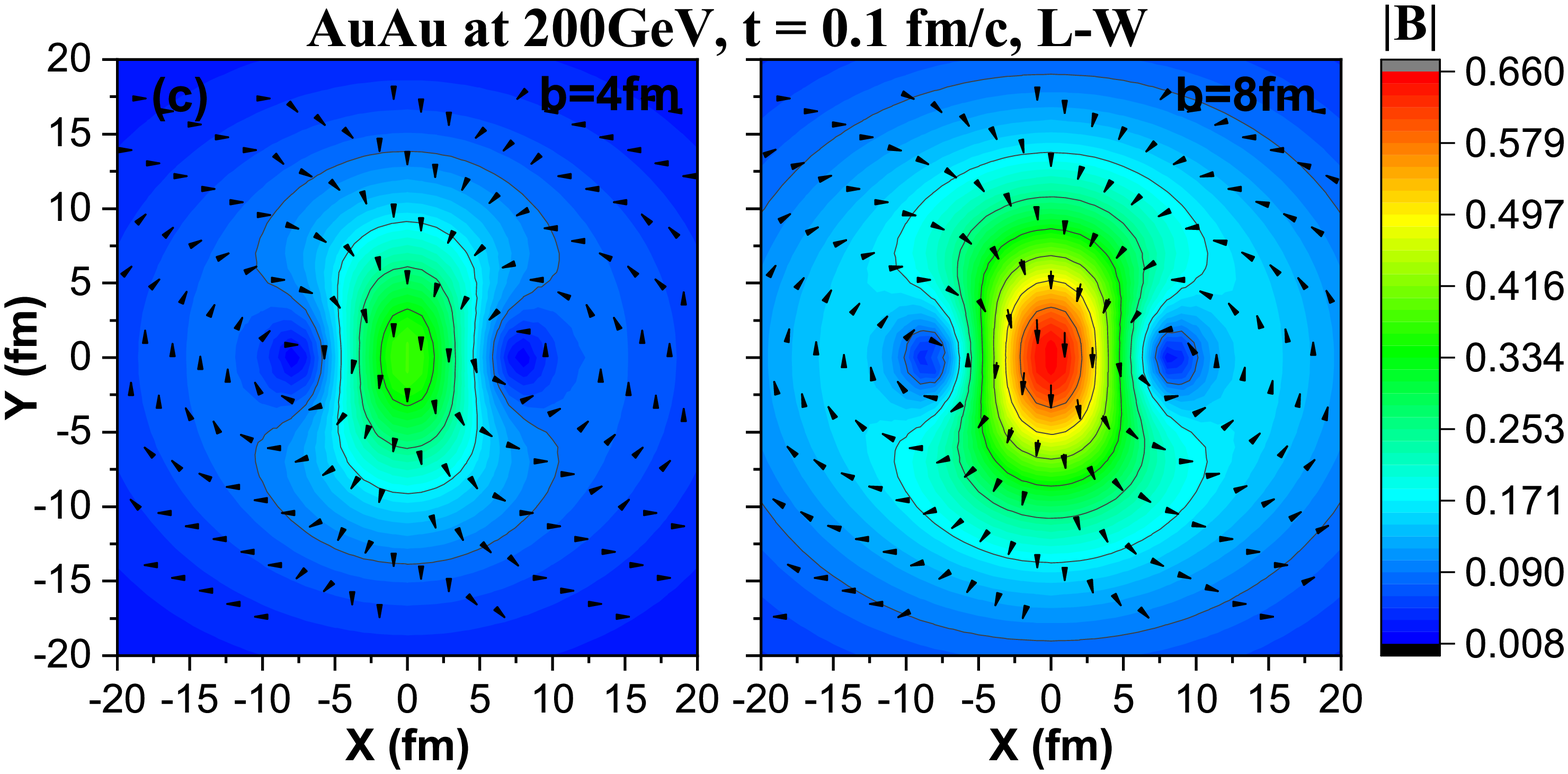} \includegraphics[width=7.7cm]{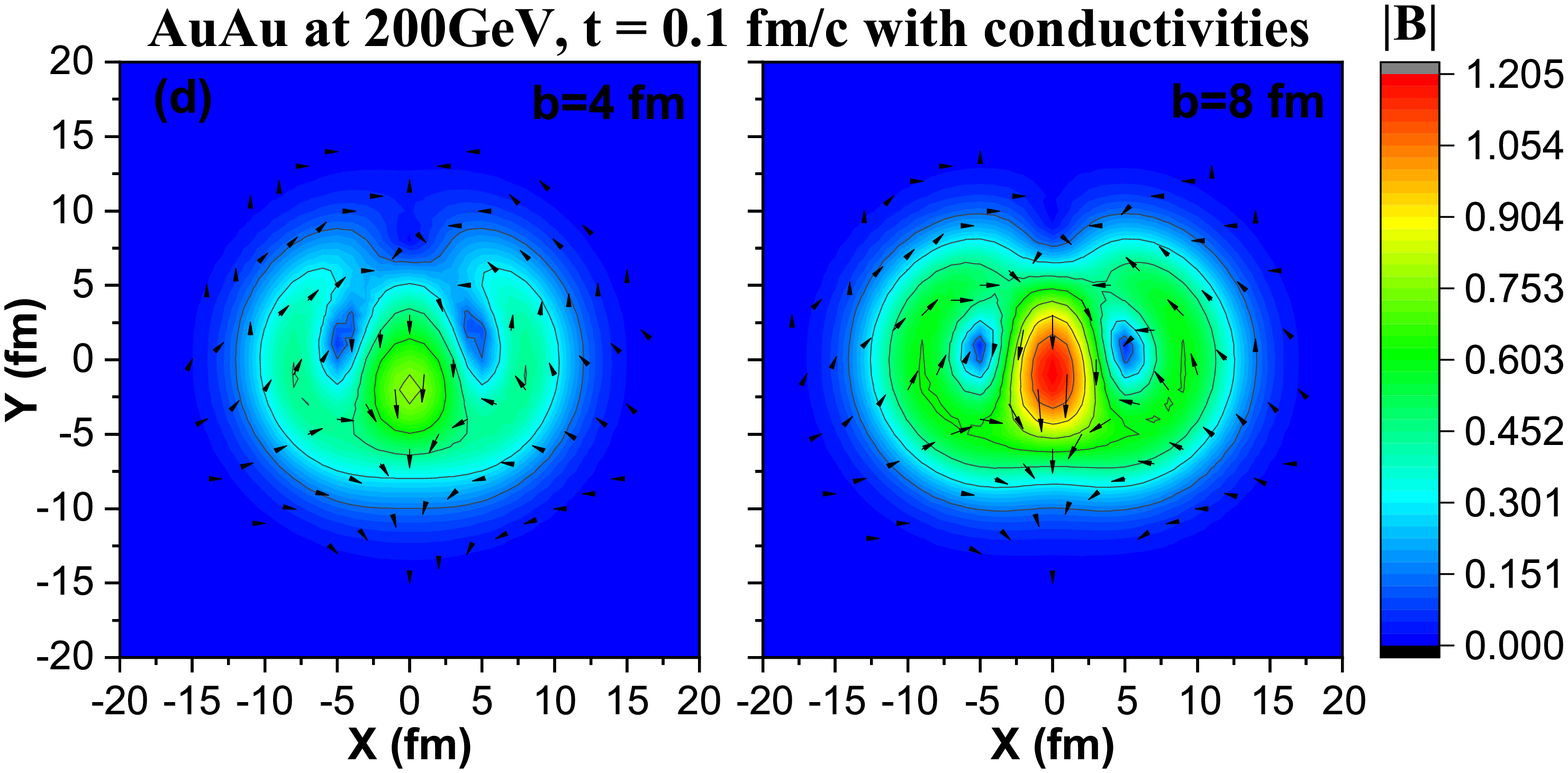} 
\par\end{centering}
\caption{\label{fig:Emod_spatial} (Color online) The spatial distributions
of the magnitude of $e\textbf{E}_{\mathrm{T}}$ (upper panels) and
$e\textbf{B}_{\mathrm{T}}$ (lower panels) (in the unit of $m_{\pi}^{2}$),
together with their two-dimensional vector fields in 200~AGeV Au+Au
collisions at $t=0.1$~fm/$c$, compared between zero \textit{vs.}
finite conductivities and different impact parameters.}
\end{figure*}


For a better illustration of the field configuration, we present
the two-dimensional vector fields of $\textbf{E}_{\mathrm{T}}$ (upper
panels) and $\textbf{B}_{\mathrm{T}}$ (lower panels) in the transverse
plane at $z=0$ in Fig.~\ref{fig:Emod_spatial}, in which the contour
plots are for the magnitude of $|\textbf{E}_{\mathrm{T}}|$ and $|\textbf{B}_{\mathrm{T}}|$.
One can observe a clear broken symmetry of both $|\textbf{E}_{\mathrm{T}}|$
and $|\textbf{B}_{\mathrm{T}}|$ about the $y=0$ axis after finite
conductivities are introduced. The breaking appears stronger for the
magnetic field than the electric field. 
Moreover, as shown by the vector field, one can clearly see the zero
electric field near the origin $(0,0,0)$, while a finite magnetic
field along $-\hat{y}$. Note that without conductivity, the magnetic
field follows the $-\hat{y}$ direction along the $x=0$ axis. However,
its direction changes after conductivities are introduced, especially
when the position is away from the origin. This would affect the inner
product between electric and magnetic fields, as will be discussed
in the following section.

\begin{figure*}
\begin{centering}
\includegraphics[width=7.7cm]{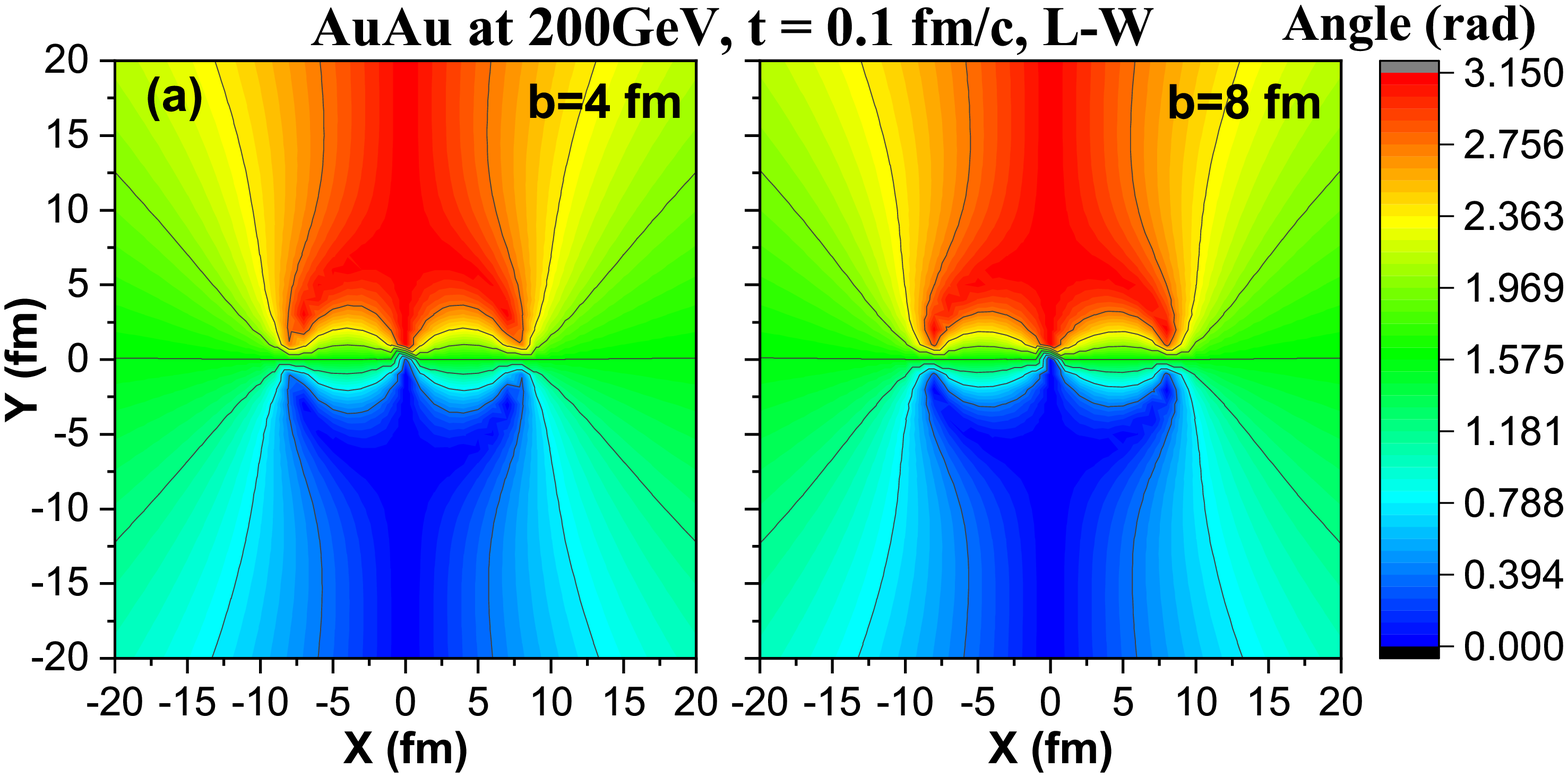} \includegraphics[width=7.7cm]{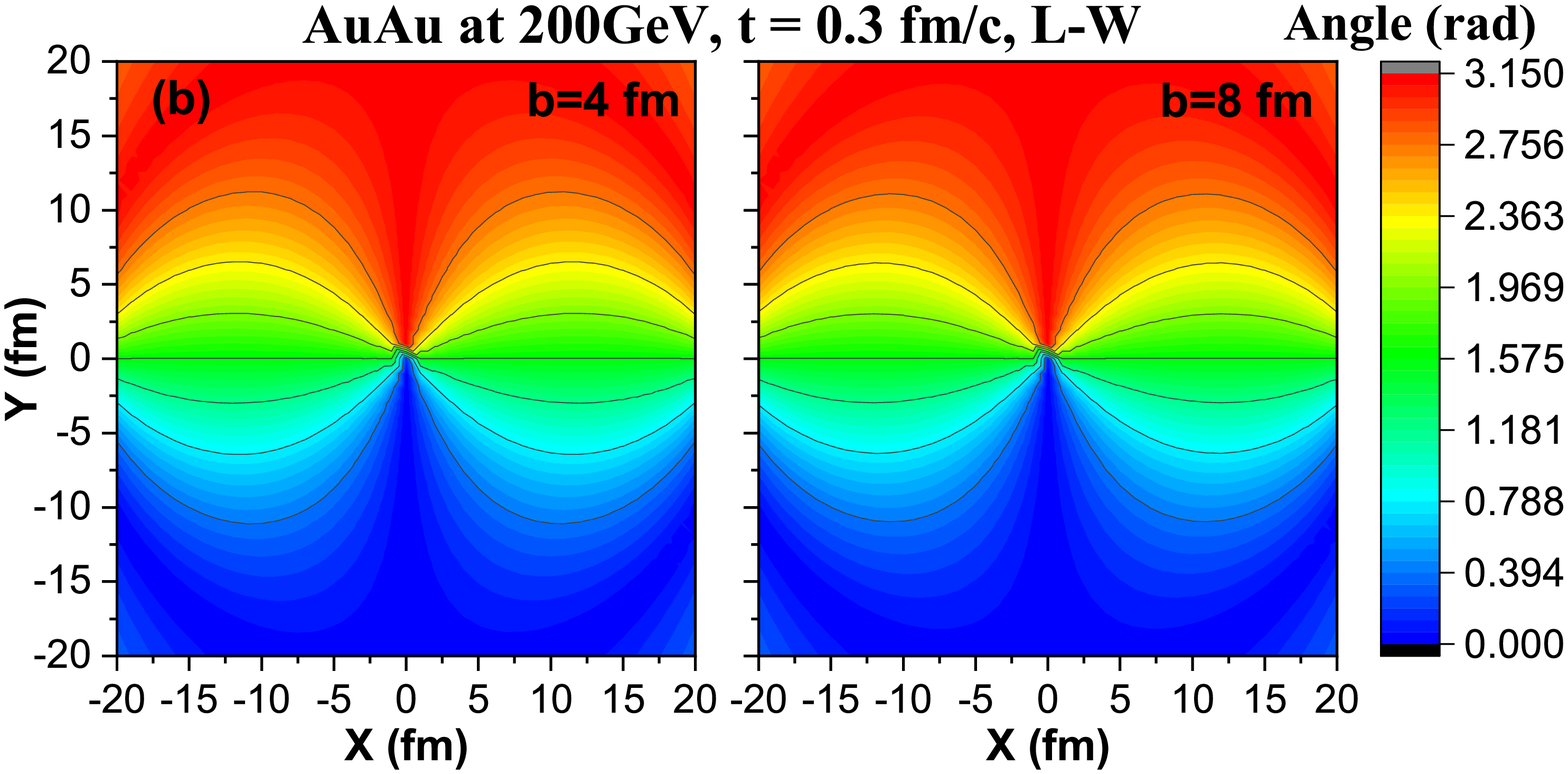} 
\par\end{centering}
\begin{centering}
\includegraphics[width=7.7cm]{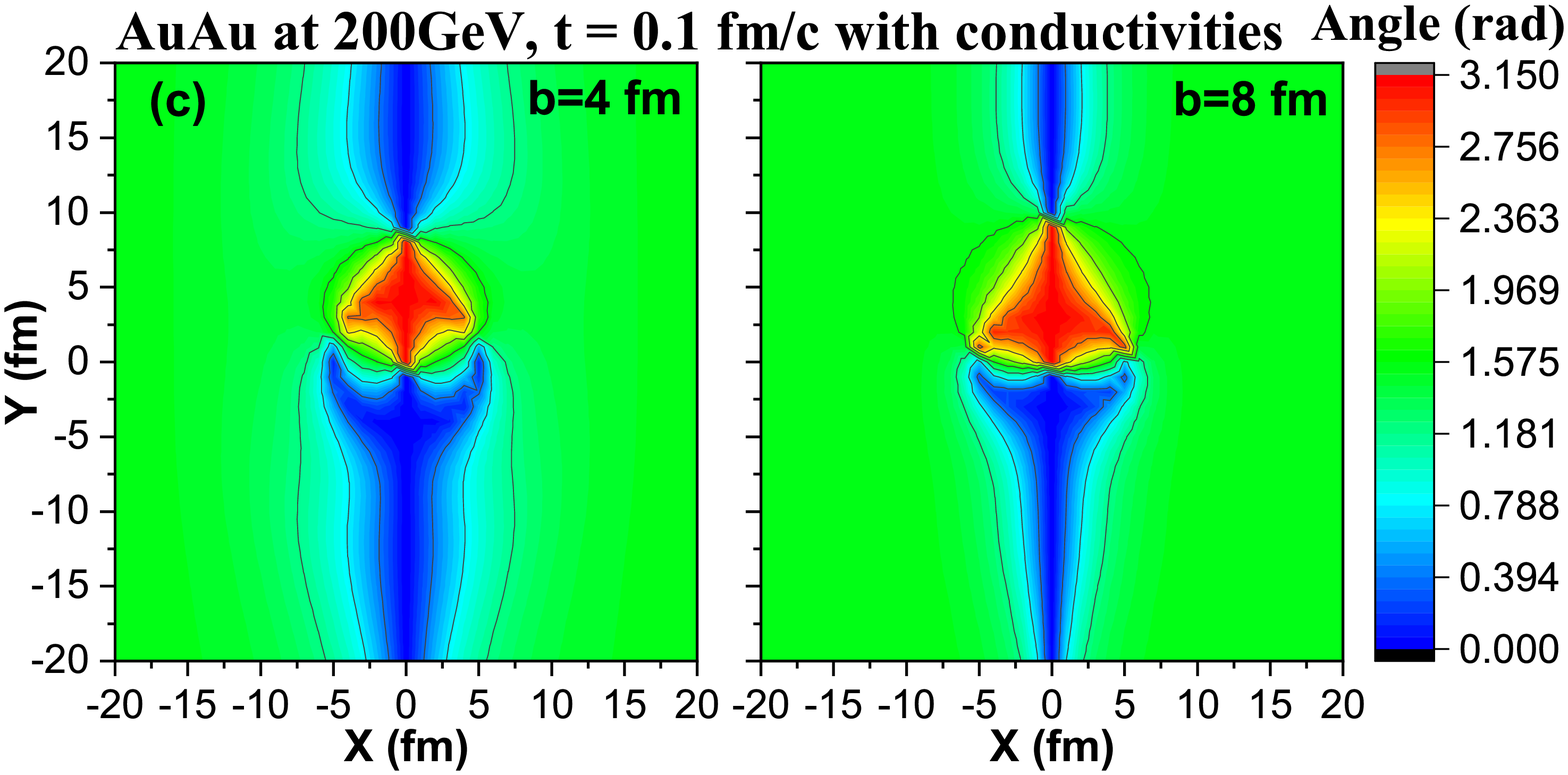} \includegraphics[width=7.7cm]{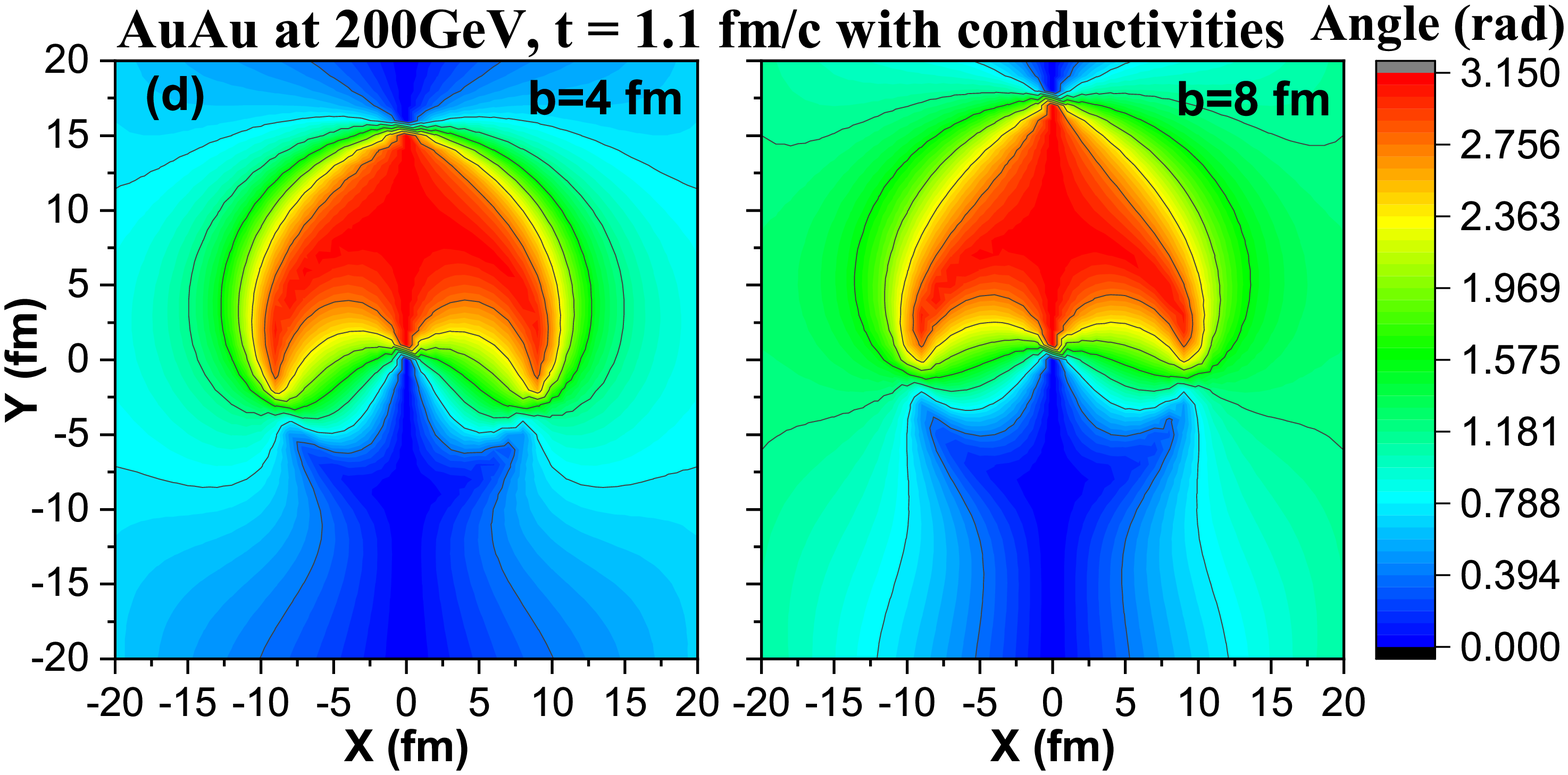} 
\par\end{centering}
\caption{\label{fig:Emod_Bmod_Angle} (Color online) The spatial distributions
of the angle between ${\mathbf{E}}_{\mathrm{T}}$ and ${\mathbf{B}}_{\mathrm{T}}$
in 200~AGeV Au+Au collisions, compared between zero \textit{vs.}
finite conductivities, different impact parameters and evolution times.}
\end{figure*}

\begin{figure*}
\begin{centering}
\includegraphics[width=7.7cm]{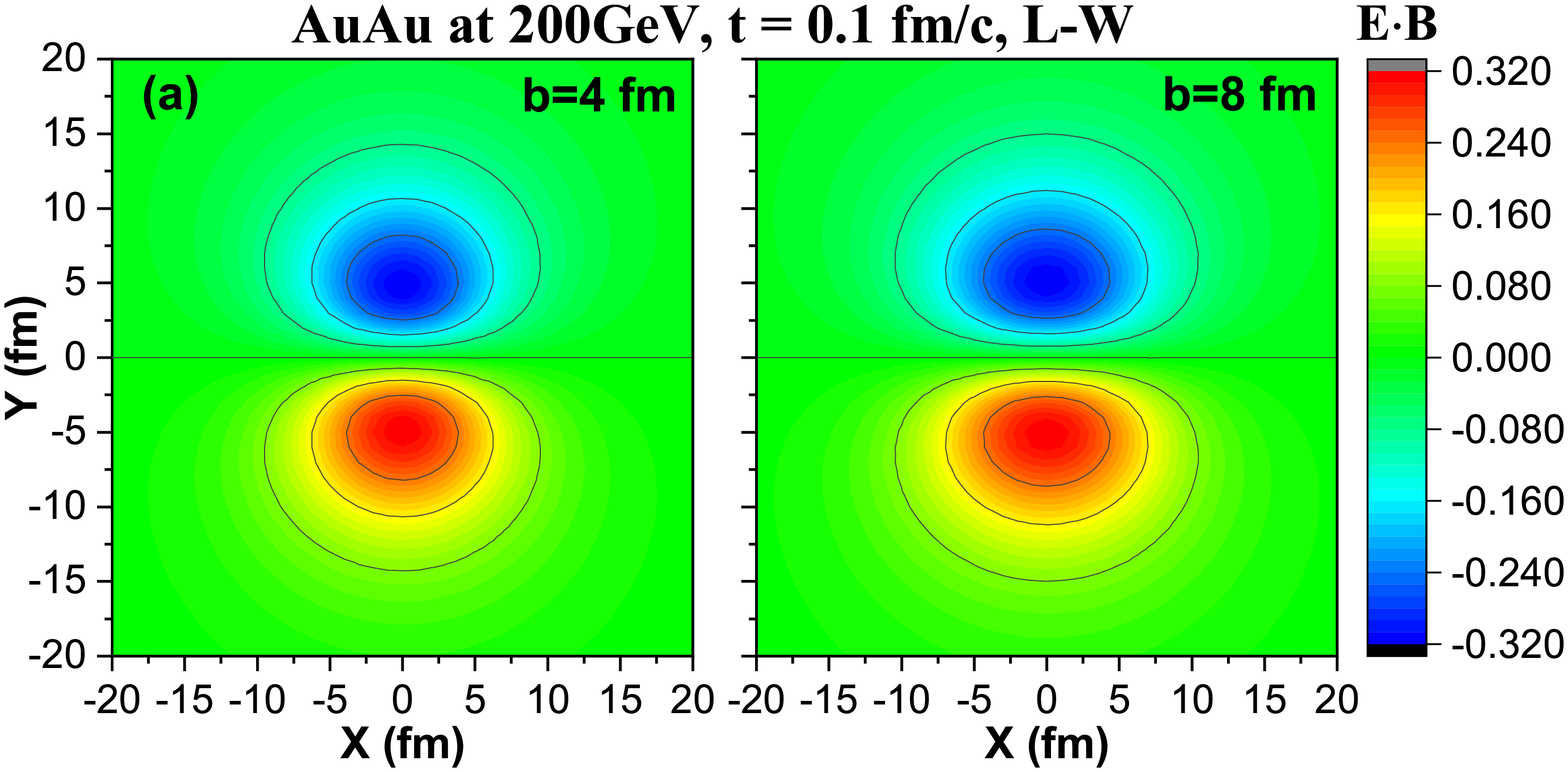} \includegraphics[width=7.7cm]{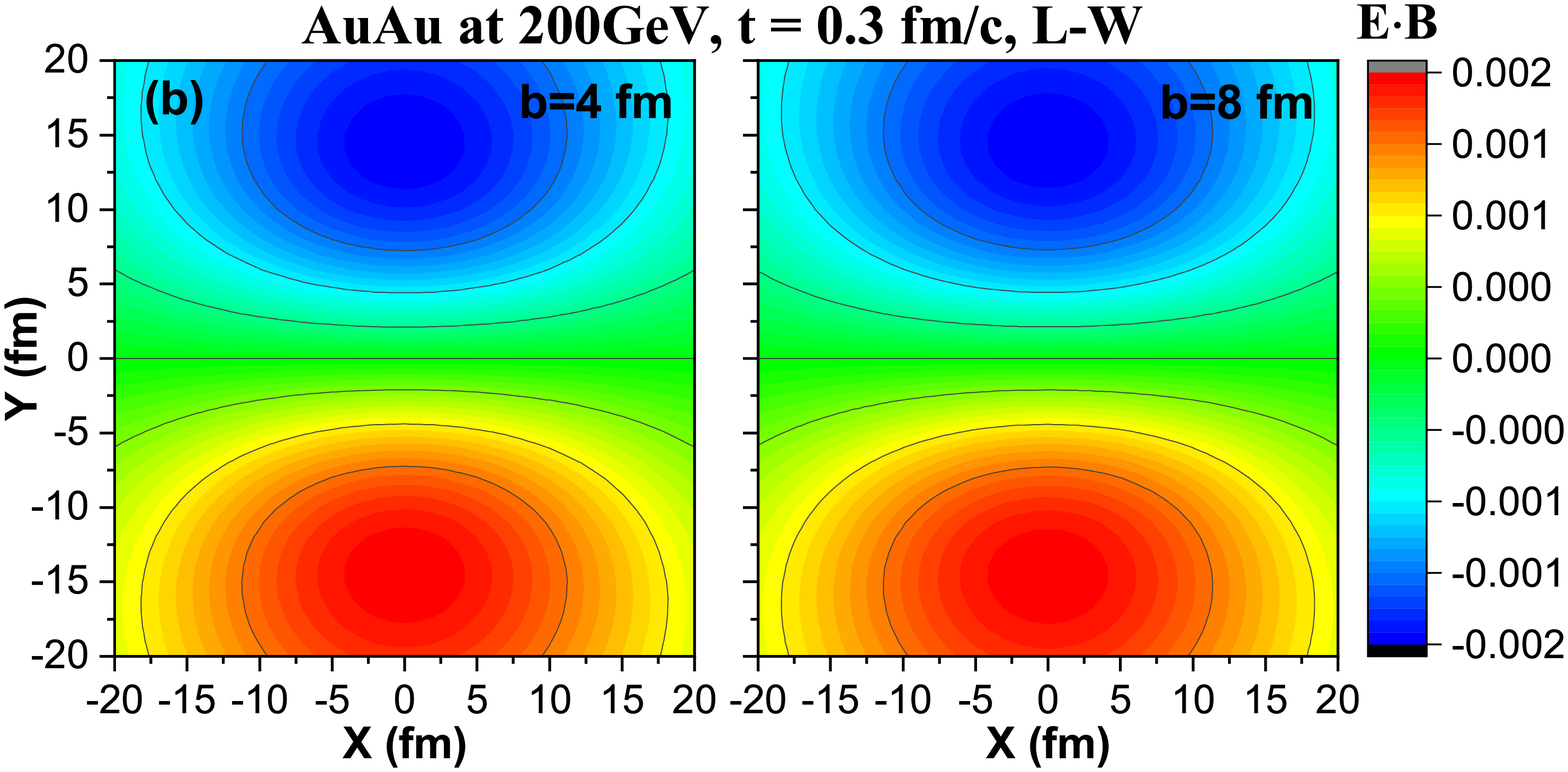} 
\par\end{centering}
\begin{centering}
\includegraphics[width=7.7cm]{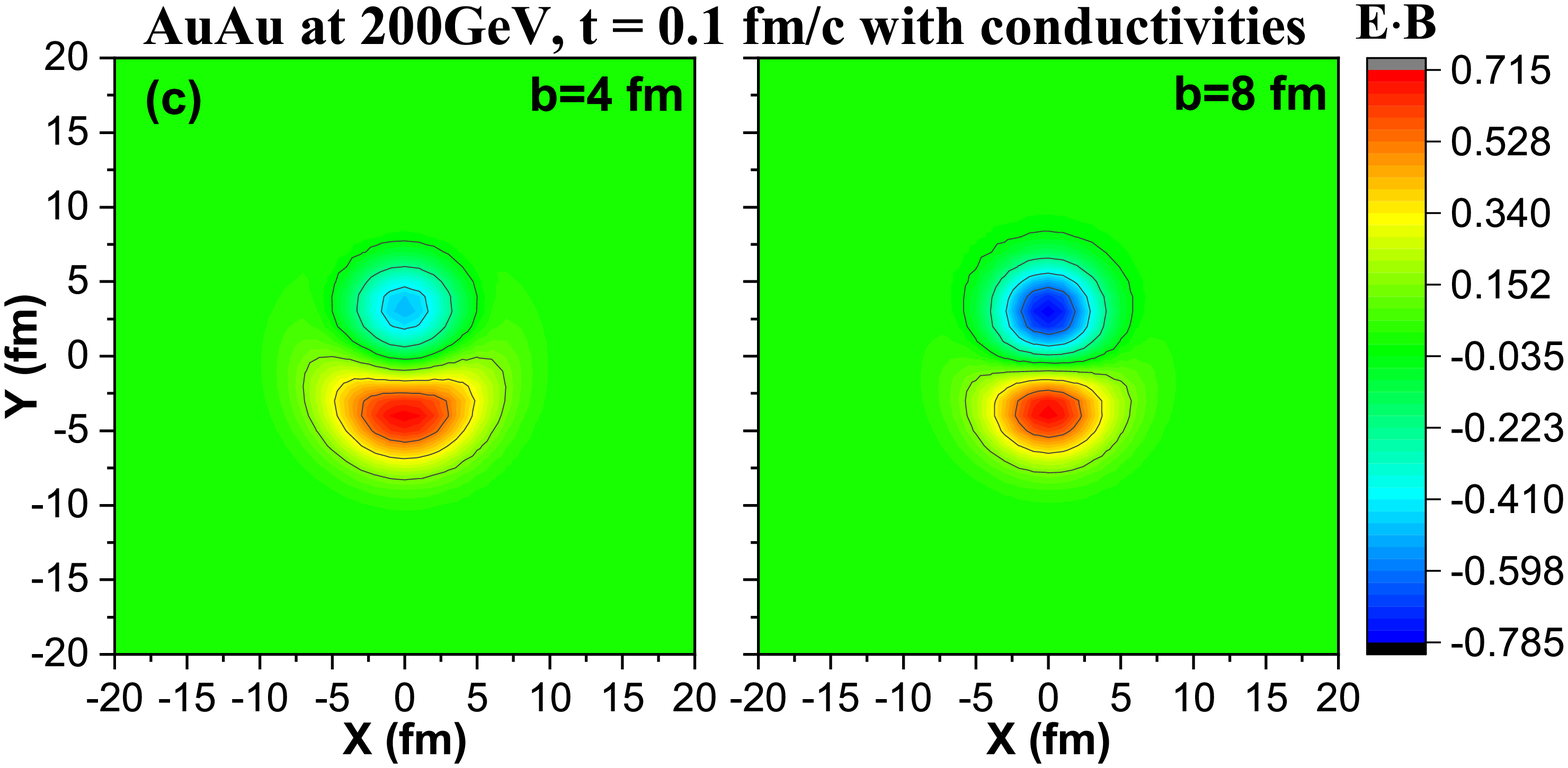} \includegraphics[width=7.7cm]{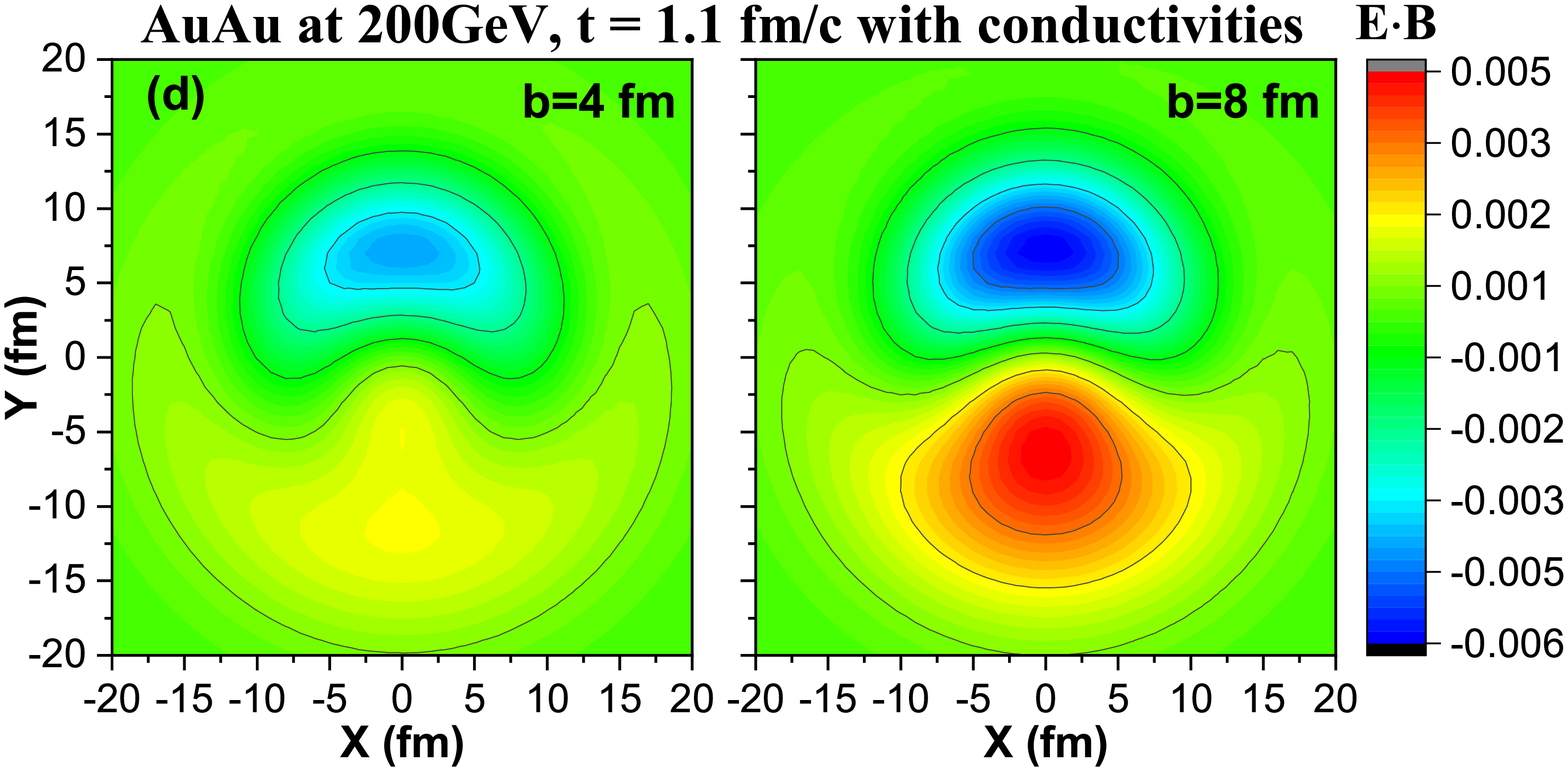} 
\par\end{centering}
\caption{\label{fig:E_dot_B_spatial} (Color online) The spatial distributions
of $e^{2}\mathbf{E}\cdot\mathbf{B}$ (in the unit of $m_{\pi}^{4}$)
in 200~AGeV Au+Au collisions, compared between zero \textit{vs.}
finite conductivities, different impact parameters and evolution times.}
\end{figure*}

\section{\label{sec:4}Spatial distributions of $\text{\textbf{E}}\cdot\boldsymbol{\text{B}}$}

With the separate results of $\mathbf{E}$ and $\mathbf{B}$ fields
above, we further investigate the spatial distribution of their inner
product ($\mathbf{E}\cdot\mathbf{B}=E_{x}B_{x}+E_{y}B_{y}+E_{z}B_{z}$),
which is directly related to the generation of the electric quadrupole
moment. We keep contributions from both transverse and longitudinal
components for the inner product for completeness, though one may
also neglect the longitudinal part~\cite{Zhao:2019ybo} due to its
relatively small contribution.



In Fig.~\ref{fig:Emod_Bmod_Angle}, we first present the spatial
distribution of the angle between electric and magnetic fields in
the transverse plane. For the most central collisions ($b=0$), one
may consider orthogonality between ${\mathbf{E}}_{\mathrm{T}}$ and
${\mathbf{B}}_{\mathrm{T}}$ due to the vanishing magnitude of the
magnetic field. At finite impact parameter, ${\mathbf{E}}_{\mathrm{T}}$
and ${\mathbf{B}}_{\mathrm{T}}$ are orthogonal to each other around
the $y=0$ axis, but appear parallel or anti-parallel around the $x=0$
axis. In vacuum, one expects to see anti-parallel alignment in the
$y>0$ half plane while parallel alignment in the $y<0$ half plane.
However, after conductivities are introduced, parallel configuration
can also be observed in the $y>0$ half plane. This is mainly due
to the direction flip of the magnetic field, as has been discussed
in Fig.~\ref{fig:Emod_spatial}. As time evolves, the pattern of
these angular distributions expand outwards in the transverse plane.
A faster expansion is seen in vacuum than in a conducting medium.

Shown in Fig.~\ref{fig:E_dot_B_spatial} is the distribution of $\mathbf{E}\cdot\mathbf{B}$,
compared between zero \textit{vs.} finite conductivities, and different
impact parameters and evolution times. One naturally expect to see
zero values for the most central collisions, 
while finite value for peripheral collisions. Symmetric distributions
are observed with respect to the reaction plane (or the $y=0$ axis)
for the Lienard-Wiechert solution, as presented in Ref.~\cite{Zhao:2019ybo}.
However, these distributions become asymmetric after finite $\sigma$
and $\sigma_{\chi}$ are included. Despite the asymmetric distribution,
one can still observe a dipole structure of $\mathbf{E}\cdot\mathbf{B}$,
i.e., opposite signs in the $y>0$ and $y<0$ half planes. The generation
of this dipole structure can be understood with the angular distributions
in Fig.~\ref{fig:Emod_Bmod_Angle}, where the electric and magnetic
fields are generally parallel to each other in the $y<0$ half plane
while anti-parallel for $y>0$. Although parallel alignment is also
seen in the $y>0$ region, the small magnitude of the fields at those
positions far away from the origin prevents the breaking of the overall
dipole structure.




\begin{figure}
\begin{centering}
\includegraphics[width=7cm]{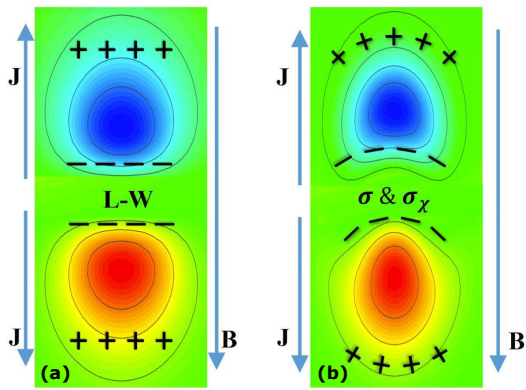} 
\par\end{centering}
\caption{\label{fig:The-coupling-of-E.B} (Color online) Generation of an electric
quadrupole moment from coupling between $\mathbf{E}\cdot\mathbf{B}$
and $\textbf{B}$ in non-central 200~AGeV Au+Au collisions, compared
between zero (left panel) and finite (right panel) conductivities.}
\end{figure}

Non-zero $\mathbf{E}\cdot\mathbf{B}$ implies non-zero $\mu_{A}$.
According to Eq.~(\ref{eq:eq0_1}), the CME current can be induced
in the presence of a magnetic field. As illustrated in Fig.~\ref{fig:The-coupling-of-E.B},
with a magnetic field aligning towards $-\hat{y}$, a negative $\mathbf{E}\cdot\mathbf{B}$
in the $y>0$ region induces a current $\mathbf{J}$ towards $+\hat{y}$,
while a positive $\mathbf{E}\cdot\mathbf{B}$ in the $y<0$ region
makes $\mathbf{J}$ along $-\hat{y}$, generating an electric quadrupole
moment in the end. This would guide positive charges into the out-of-plane
direction of heavy-ion collisions, while negative charges into the
in-plane direction, giving rise to the charge separation of hadron
$v_{2}$ even without the formation of CMW. Taking into account the
finite $\sigma$ and $\sigma_{\chi}$ affects the direction of the
local CME current, resulting in a quantitatively different electric
quadrupole moment compared to the vacuum scenario, although the qualitative
picture is still consistent with the findings proposed in Ref.~\cite{Zhao:2019ybo}.


\begin{figure}
\begin{centering}
\includegraphics[width=7cm]{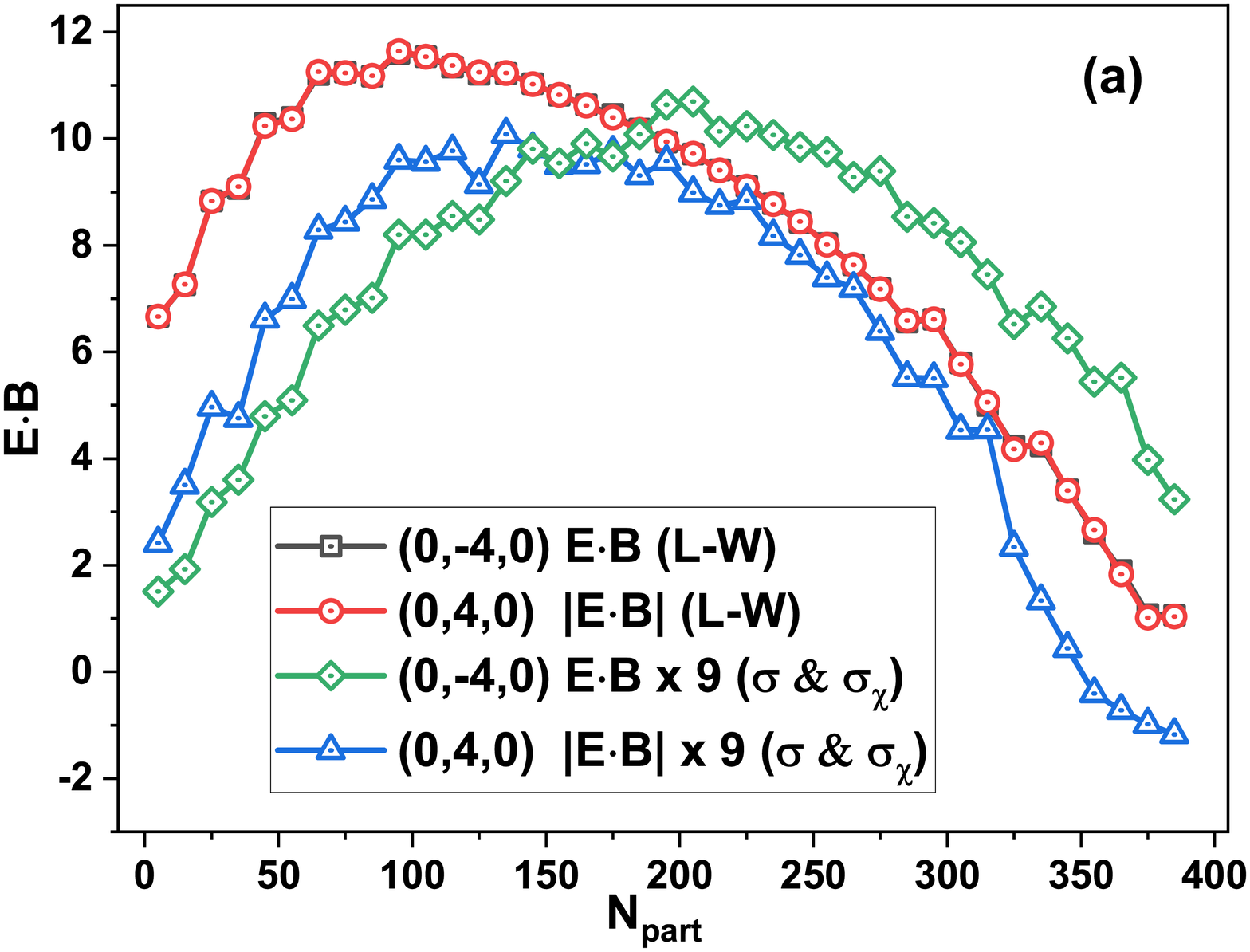}\hspace{0.2cm}\includegraphics[width=7.2cm]{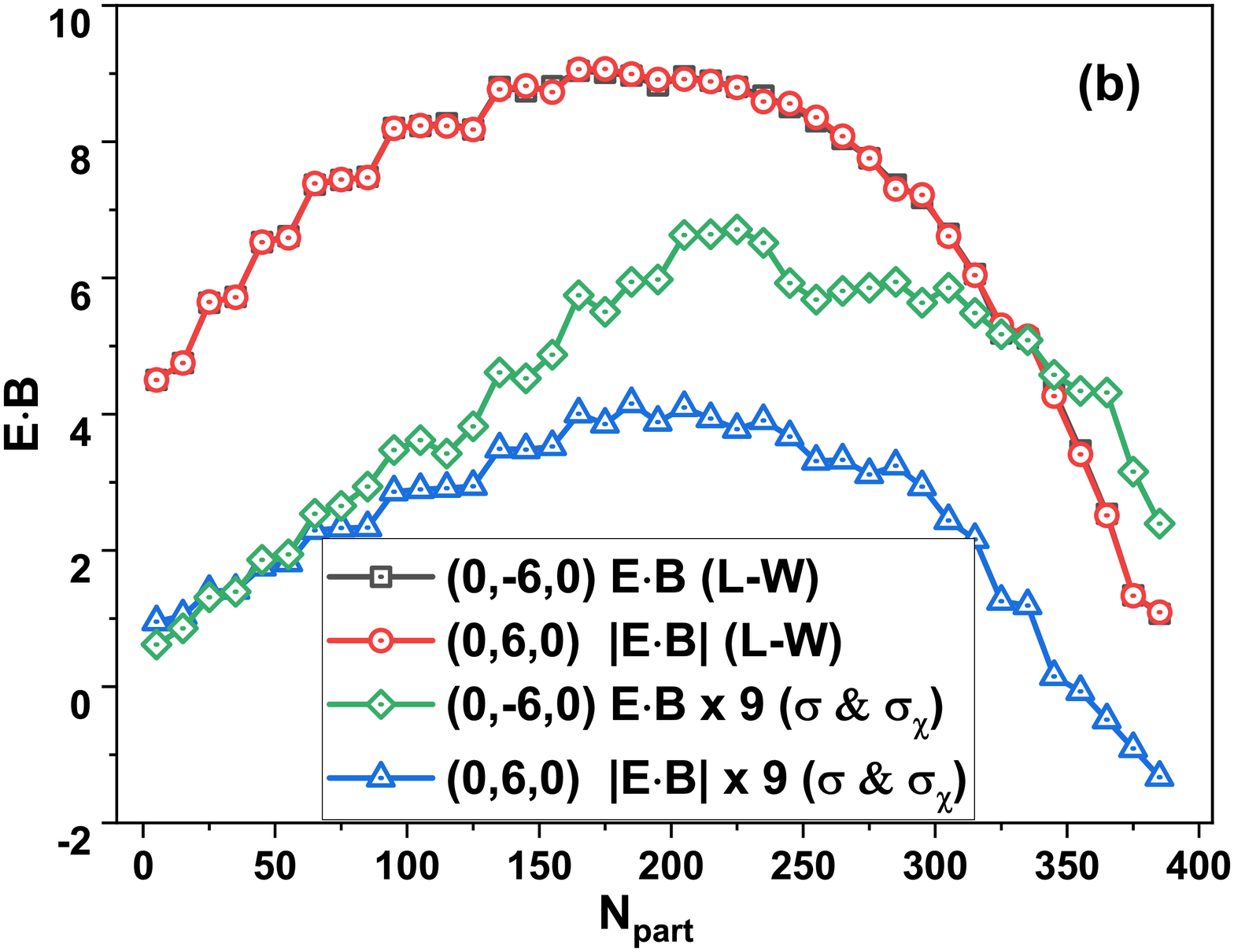} 
\par\end{centering}
\caption{\label{E.B040npart} (Color online) The value of $e^{2}\textbf{E}\cdot\textbf{B}$
(in the unit of $m_{\pi}^{4}$) as a function of $N_{\mathrm{part}}$
at the initial time, and positions of $\textbf{r}=(0,\pm4~\mathrm{fm},0)$
(upper panel) and $\textbf{r}=(0,\pm6~\mathrm{fm},0)$ (lower panel)
in 200~AGeV Au+Au collisions, compared between zero and finite conductivities.}
\end{figure}

\begin{figure}
\begin{centering}
\includegraphics[width=7cm,height=5.2cm]{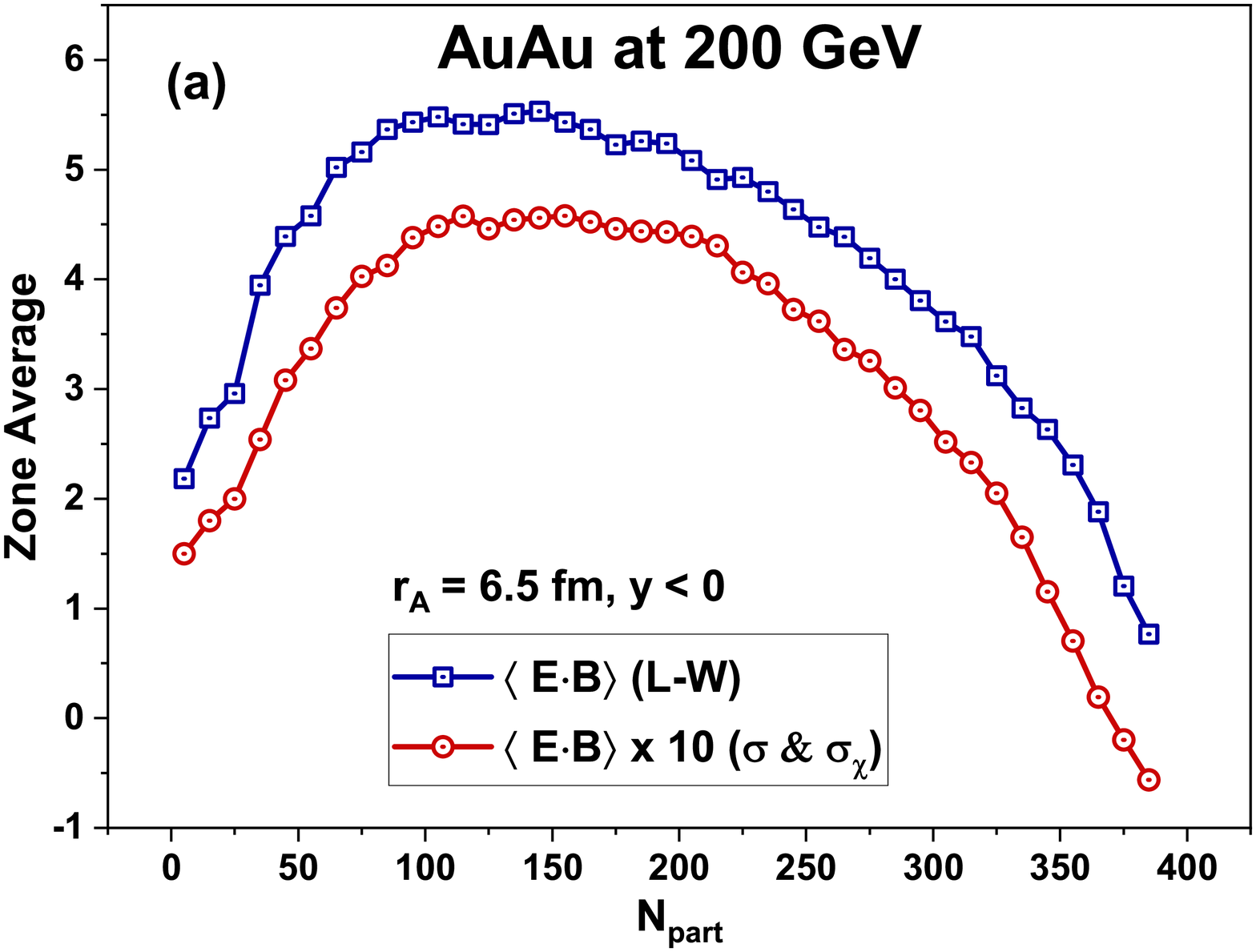}\hspace{0.2cm}\includegraphics[width=7.7cm]{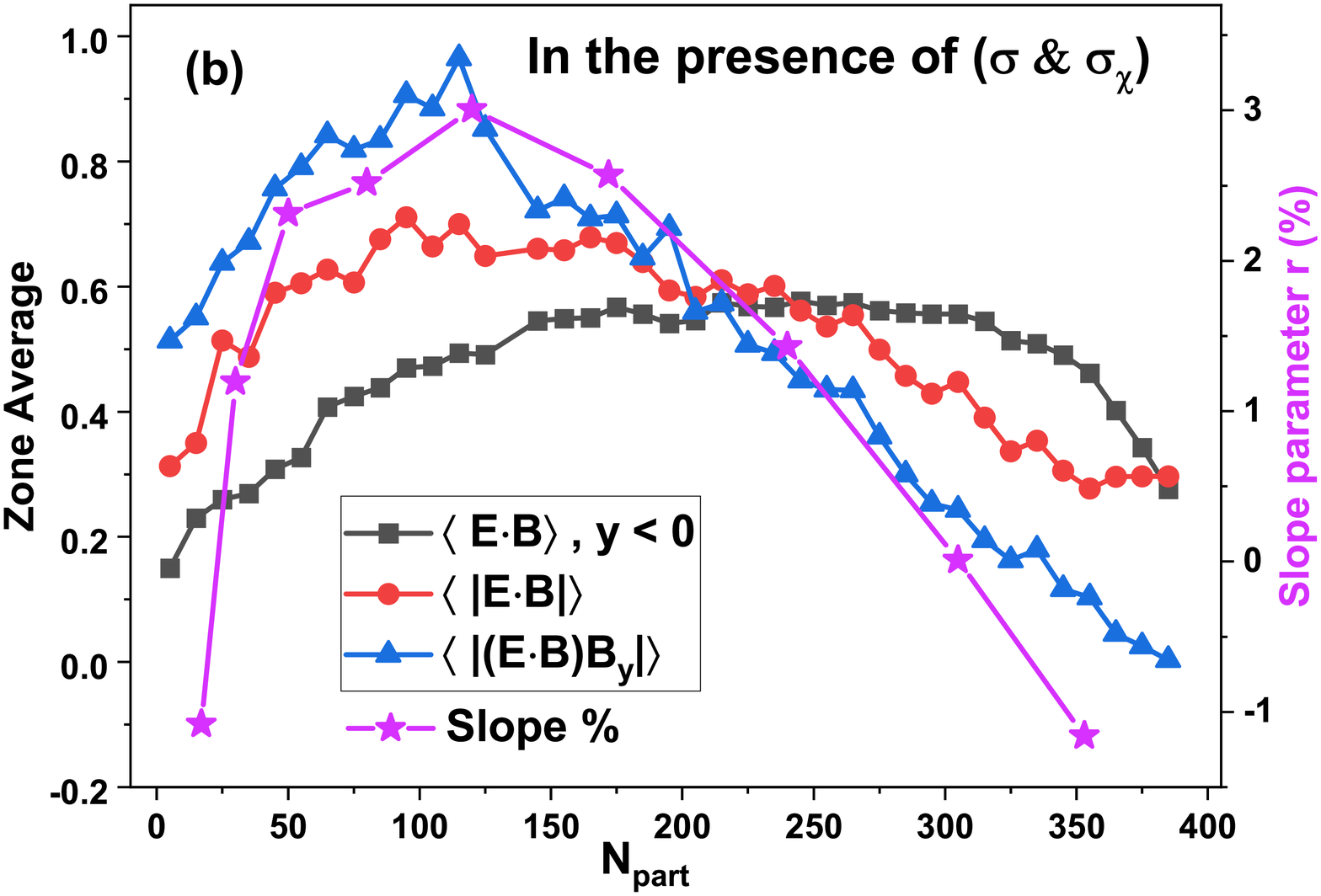}
\par\end{centering}
\caption{\label{fig:The-zone-averaged-density} (Color online) The zone-averaged
$|{\textbf{E}}\cdot{\mathbf{B}}|$ (in the unit of $m_{\pi}^{4}$)
and $|({\textbf{E}}\cdot{\mathbf{B}})B_{y}|$ (in the unit of $m_{\pi}^{6}$)
in the geometric overlapping region between colliding nuclei at the
initial time of 200~AGeV Au+Au collisions, as a function of $N_{\mathrm{part}}$,
compared between zero and finite conductivities (upper panel), and
between different average schemes and the slope parameter $r$ measured
by the STAR Collaboration~\cite{STAR:2015wza} (lower panel).}
\end{figure}

To further investigate how the conductivities quantitatively affect
the electric quadrupole moment, in Fig.~\ref{E.B040npart} we compare
the participant number ($N_{\mathrm{part}}$) dependence of $\mathbf{E}\cdot\mathbf{B}$
between the Lienard-Wiechert solution and the solution of the Maxwell
equations with finite conductivities. Results are shown for different
locations at the initial time. In the upper panel, we observe that
at $(0,-4~\mathrm{fm},0)$, $\mathbf{E}\cdot\mathbf{B}$ from with
and without conductivities share similar shape of the $N_{\mathrm{part}}$
dependence. It first increases and then decreases as $N_{\mathrm{part}}$
increases, since the electric field is small at large impact parameter
(small $N_{\mathrm{part}}$) while the magnetic field is small at
small impact parameter (large $N_{\mathrm{part}}$). On the other
hand, the magnitude of $\mathbf{E}\cdot\mathbf{B}$ with conductivities
is about 9 times smaller than that without conductivities at the initial
time. In addition, while the absolute value of $\mathbf{E}\cdot\mathbf{B}$
with zero conductivity are symmetric at $(0,-4~\mathrm{fm},0)$ and
$(0,+4~\mathrm{fm},0)$ (the black and red curves overlap each other),
such symmetry is broken (between the blue and green curves) after
finite conductivities are introduced. Similar findings have also been
confirmed in the lower panel for the locations of $(0,\pm6~\mathrm{fm},0)$.

Shown in Fig.~\ref{fig:The-zone-averaged-density} is the zone averaged
$|\mathbf{E}\cdot\mathbf{B}|$ as a function of the participant number
at the initial time. The average is conducted over the geometric overlapping
region between the two colliding nuclei, i.e., region that simultaneously
satisfies $(x-b/2)^{2}+y^{2}<r_{\mathrm{A}}^{2}$ and $(x+b/2)^{2}+y^{2}<r_{\mathrm{A}}^{2}$,
with $b$ being the impact parameter and $r_{\mathrm{A}}$ being the
nucleus radius parameter taken as 6.5~fm here. Uncertainties from
taking different $r_{\mathrm{A}}$ values and applying different average
schemes have been discussed in Ref.~\cite{Zhao:2019ybo} and found
small. In the upper panel of Fig.~\ref{fig:The-zone-averaged-density},
we first follow Ref.~\cite{Zhao:2019ybo} to present the zone averaged
value of $\mathbf{E}\cdot\mathbf{B}$ in the $y<0$ half plane. Similar
to previous results at a specific location, the zone averaged $\mathbf{E}\cdot\mathbf{B}$
share a similar shape with respect to $N_{\mathrm{part}}$ between
zero and finite conductivities, although the magnitude at the initial
time becomes much smaller after conductivities are included.

Since the electromagnetic field in a conducting medium is asymmetric
about the reaction plane, averaging in the $y<0$ half plane is no
longer a good representation of the dipole structure of $\mathbf{E}\cdot\mathbf{B}$
over the whole overlapping region. Therefore, in the lower panel of
Fig.~\ref{fig:The-zone-averaged-density}, we compare different average
schemes for the finite conductivity scenario. Visible difference can
be observed between averaging $\mathbf{E}\cdot\mathbf{B}$ over the
$y<0$ half plane (black curve) and averaging $|\mathbf{E}\cdot\mathbf{B}|$
over the whole overlapping region (red curve). 
In this lower panel, the shape of the average $|\mathbf{E}\cdot\mathbf{B}|$
is also compared to the slope parameter of the charge separation of
the hadron $v_{2}$ measured by the STAR Collaboration~\cite{STAR:2015wza}
(purple), as proposed in Ref.~\cite{Zhao:2019ybo}. The slope parameter
$r$, defined via $v_{2}(\pi^{\pm})=v_{2}^{\mathrm{base}}(\pi^{\pm})\mp rA_{\mathrm{ch}}/2$
with the charge asymmetry of the collision system given by $A_{\mathrm{ch}}=(N_{+}-N_{-})/(N_{+}+N_{-})$,
quantifies the different $v_{2}$ between $\pi^{+}$ and $\pi^{-}$.
As shown in the figure, the average $|\mathbf{E}\cdot\mathbf{B}|$
shares a similar $N_{\mathrm{part}}$ dependence to the measured $r$
parameter, implying the QED anomaly ($\mathbf{E}\cdot\mathbf{B}$)
could be a possible source for the separation of $v_{2}$ between
positive and negative charges. Since the electric quadrupole moment
is a more direct cause of the charge separation of $v_{2}$, we also
present the zone average of $|(\mathbf{E}\cdot\mathbf{B})B_{y}|$
in the figure (blue curve). Indeed, a better qualitative agreement
is obtained with the shape of the measured $r$ parameter. Nevertheless,
a quantitative description of the experimental data would require
coupling the electromagnetic field with the QGP expansion (e.g. the
hydrodynamic model). This is beyond the scope of the present work
and will be left for a future exploration.



\begin{figure}
\begin{centering}
\includegraphics[width=7cm]{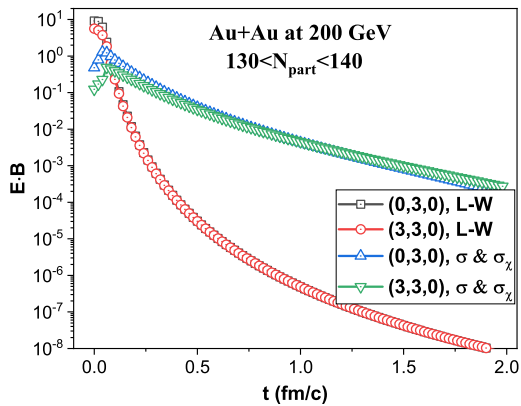} 
\par\end{centering}
\caption{\label{fig:Time-evol_E_B} (Color online) The time evolution of $e^{2}\text{\textbf{E}}\cdot\boldsymbol{\mathbf{B}}$
(in the unit of $m_{\pi}^{4}$) in 200~AGeV Au+Au collisions with
$130<N_{\mathrm{part}}<140$, compared between zero and finite conductivities,
and two different locations.}
\end{figure}

In the end, we study the time evolution of $\mathbf{E}\cdot\mathbf{B}$
in Fig.~\ref{fig:Time-evol_E_B}. Events with participant number
between 130 and 140 are selected here (corresponding to an impact
parameter around 8~fm) for Au+Au collisions at $\sqrt{s_{\mathrm{NN}}}=200$~GeV.
Results at two different locations, $(0,3~\mathrm{fm},0)$ and $(3~\mathrm{fm},3~\mathrm{fm},0)$,
are presented and compared between zero and finite conductivity scenarios.
One can observe although the zero conductivity scenario starts with
a larger $\mathbf{E}\cdot\mathbf{B}$ than the finite conductivity
scenario, as has also been observed previously in Figs.~\ref{E.B040npart}
and~\ref{fig:The-zone-averaged-density}, the former decays much
faster than the latter. Therefore, including finite $\sigma$ and
$\sigma_{\chi}$ helps extend the influence of the electromagnetic
field to a much later evolution stage of the QGP. Since stronger elliptic
flow of the medium will be developed towards later time, introducing
the electric and chiral magnetic conductivities may also quantitatively
enhance the charge separation of $v_{2}$, or the slope parameter
$r$.

\section{Summary and outlook}

\label{sec:summary}

In this work, we have conducted a systematic study on the effects
of the electric ($\sigma$) and chiral magnetic ($\sigma_{\chi}$)
conductivities on the spacetime evolution of the electromagnetic fields
generated in high-energy nuclear collisions. By coupling the charge
distribution from a MC Glauber model with the solution of the Maxwell
equations that include both $\sigma$ and $\sigma_{\chi}$, or its
zero conductivity limit (Lienard-Wiechert), we have calculated the
time evolution of the spatial distributions of electric ($\textbf{E}$)
and magnetic ($\textbf{B}$) fields, together with the electromagnetic
anomaly ($\textbf{E}\cdot\textbf{B}$) and the electric quadrupole
moment ($(\textbf{E}\cdot\textbf{B})\textbf{B}$) at both zero and
finite conductivities.

Our results show that although the electromagnetic field in vacuum
is about an order of magnitude stronger than that in a conducting
medium at the initial time, the former decays much faster than the
latter. Additionally, in the transverse plane, while $|\textbf{E}|$
and $|\textbf{B}|$ appear symmetric about both $x=0$ and $y=0$
axes at zero conductivities, a broken symmetry about the
$y=0$ axis, or the reaction plane, is observed after finite conductivities
are introduced. This symmetry breaking is mainly from the non-vanishing
azimuthal component with the presence of $\sigma_{\chi}$ for the
electric field $\textbf{E}_{\mathrm{T}}$, while from the non-vanishing
radial component with the presence of $\sigma_{\chi}$ for the magnetic
field $\textbf{B}_{\mathrm{T}}$. The magnitudes of the longitudinal
components of both $\textbf{E}$ and $\textbf{B}$ appear much smaller
than their transverse component, while no symmetry breaking is observed
for $E_{z}$ and $B_{z}$ after finite conductivities are introduced.
A clear dipole structure for $\textbf{E}\cdot\textbf{B}$ and a quadrupole
pattern for $(\textbf{E}\cdot\textbf{B})\textbf{B}$ are still observed
in our results although they are both distorted compared to the vacuum
scenario due to the symmetry breaking of $\textbf{E}$ and $\textbf{B}$
fields in a conducting medium. Since the magnitude of $\textbf{E}$
decreases, while the magnitude of $\textbf{B}$ increases as the impact
parameter increases, one can observe a non-monotonic dependence (first
increase and then decrease) of $\textbf{E}\cdot\textbf{B}$ and $(\textbf{E}\cdot\textbf{B})\textbf{B}$
with respect to the nucleon participant number in heavy-ion collisions.
These dependences are found qualitatively consistent with the STAR
data on the slope parameter $r$ as a function of the participant
number, indicating the QED anomaly could be an underlying mechanism
that drives the $v_{2}$ separation between positive and negative
charges. Since $(\textbf{E}\cdot\textbf{B})\textbf{B}$ is more directly
related to the electric quadrupole moment that gives rise to the charge
separation, it appears to agree with the experimental data better
than $\textbf{E}\cdot\textbf{B}$.

While this work provides a more quantitative understanding of the
spacetime evolution of electromagnetic field and electromagnetic anomaly
in relativistic heavy-ion collisions, it should be further improved
in several directions. For instance, it is necessary to couple these
profiles of electromagnetic field to hydrodynamic models or transport
models for a more direct comparison to the charged particle observables,
from which one may draw more solid conclusion about whether the QED
anomaly is the key mechanism of the charge separation of the hadron
$v_{2}$. In addition, we assumed constant values of $\sigma$ and
$\sigma_{\chi}$ in the present study, which should vary
as the QGP expands. Last but not least, apart from the slope parameter
$r$ of the $v_{2}$ separation, there exist other observables that
may help place more stringent constraints on the electromagnetic field
inside a conducting medium, such as the directed flow coefficient
($v_{1}$) of heavy quarks, whose precise theoretical description
still remains a challenge with simplified modelings of the electromagnetic
field in literature. We will extend our study to these aspects in
our upcoming efforts.

\begin{acknowledgments}
We are grateful to Xin-Li Sheng, Qun Wang and Xin-Li Zhao for very
helpful discussions. This work was supported by the National Natural
Science Foundation of China (NSFC) under Grant Nos. 12175122 and 2021-867. 
\end{acknowledgments}

\bibliographystyle{unsrt}
\phantomsection\addcontentsline{toc}{section}{\refname}\bibliography{SCrefs}

\end{document}